\documentclass[useAMS,usenatbib]{mn2e}

\pdfoutput=1

\usepackage{amssymb}
\usepackage{amsmath}
\usepackage{graphicx}
\usepackage{epstopdf}

\newcommand{\vect}[1]{\boldsymbol{#1}}

\title[Using the HEX Technique for Approximating Simulated Haloes]{A Halo Expansion (HEX) Technique for Approximating Simulated Dark Matter Haloes}
\author[Ben Lowing et al]{Ben Lowing$^{1}$\thanks{E-mail: b.j.lowing@durham.ac.uk}, Adrian Jenkins$^{1}$, Vincent Eke$^{1}$ and Carlos Frenk$^{1}$ \\
$^{1}$Institute for Computational Cosmology, Department of Physics, University of Durham, South Road, Durham, DH1 3LE, UK}

\begin{document}

\date{Accepted ... Received ... in original form ...}
\pagerange{\pageref{firstpage}--\pageref{lastpage}} \pubyear{2011}
\maketitle

%%%%%%%%%%%%%%%%%%%%%%%%%%%%%%%%%%%%%%%%%%%%%%%%

\label{firstpage}

\begin{abstract}
We apply a basis function expansion method to create a
time-evolving density/potential approximation of the late
growth of simulated N-body dark matter haloes. We demonstrate how the
potential of a halo from the Aquarius Project can be accurately
represented by a small number of basis functions, and show that the halo expansion (HEX) method provides a way to replay simulations. We explore the level
of accuracy of the technique as well as some of its limitations.  We
find that the number of terms included in the expansion must be large
enough to resolve the large-scale distribution and shape of the halo
but, beyond this, additional terms result in little further
improvement. Particle and subhalo orbits can be integrated in this
realistic, time-varying halo potential approximation, at much lower
cost than the original simulation, with high fidelity for many
individual orbits, and a good match to the distributions of orbital
energy and angular momentum. Statistically, the evolution of
structural subhalo properties, such as mass, half-mass radius and
characteristic circular velocity, are very well reproduced in the halo expansion
approximation over several gigayears. We demonstrate an application of
the technique by following the evolution of an orbiting subhalo at much
higher resolution than can be achieved in the original simulation. Our
method represents a significant improvement over commonly used
techniques based on static analytical descriptions of the halo
potential.

\end{abstract}


\section{Introduction}

In the standard cosmological paradigm of structure formation
($\Lambda$CDM), dark matter haloes are built up through the repeated
hierarchical merging of smaller haloes \citep{White:1978,
Frenk:1985}. These haloes provide the sites in which galaxies
form. Any model of galaxy formation, be it an SPH simulation or a
semi-analytical calculation, must include a description of the
evolution of the halo in which the galaxy grows. These descriptions
usually take the form of either N-body simulations, analytical potential
profiles, or statistical merger trees. In this paper, we present a new
way of characterising the evolution of dark matter haloes that can be
employed in galaxy formation models, or to explore their small-scale
structure. 

\par

Nearly all representations of haloes are motivated by cosmological
N-body simulations. These are a powerful tool and have allowed us to
gain insight into the non-linear stages of halo growth. The initial
power spectrum of density fluctuations in the CDM cosmogony has power
on all scales and this affects the internal evolution of halos on a
wide range of scales. However, investigating the structure and
substructure of halos requires simulations of ever increasing
resolution and ever increasing computational expense. The
state-of-the-art are the Aquarius simulations of galactic dark matter
halos, the largest of which achieved a resolution of $\sim 10^3
M_{\odot}$ \citep{Springel:2008a}. From these and other
simulations \citep{Stadel:2009} we have learnt not only about the basic structure of
haloes - that they have approximately universal density profiles well
described by an NFW profile \citep{Navarro:1996, Navarro:1997} or that
they are strongly triaxial in shape \citep{Allgood:2006, Bett:2007, Hayashi:2007} - but also about the
properties of their small-scale structure \citep{Springel:2008a, Diemand:2008, Vogelsberger:2011}. 

\par

In spite of their impressive resolution, recent simulations have
a number of limitations. Firstly, only a few examples have been
calculated so far; secondly, their resolution is still below that
required to follow the evolution of the smallest subhaloes, including
those that host the ultrafaint dwarfs of the Milky Way; finally, they
neglect the effects of baryons in the evolution of the main halo and
its subhaloes. 
%The use of simulations is limited by the high computational cost of
%running a full N-body simulation, thus restricting it to only a few
%realisations of haloes. In order to study the origin and behaviour of
%substructure or to explore galaxy formation models, it is necessary to
%perform simulations from a wide range of initial conditions and
%including different physics. This can be done at limited resolution
%using resimulations \citep{Kazantzidis:2004c, Okamoto:2005,
%Scannapieco:2008}, but to model fully the dynamics and evolution of
%structure and objects embedded in a larger halo it is often necessary
%to resimulate the entire halo. Unfortunately, this can be
%prohibitively costly. 

The high cost of full simulations can be avoided by introducing
approximations. A commonly used one is to assume a static analytical
potential to represent the halo and then perform a live simulation of
just the small-scale component of interest. Computational resources can
then be targeted at that component and large numbers of resimulations
performed. This method has been applied to a wide range of problems
such as the orbits and evolution of subhaloes
\citep{Taylor:2001, Zentner:2003, Penarrubia:2005}, the build-up of 
galactic stellar haloes \citep{Bullock:2005}, the formation of streams
\citep{Penarrubia:2006}, or the disruption and heating of disks
\citep{Benson:2004}. 

Using an analytical potential allows the parameters of the dark matter
halo to be varied in a way that cannot be done in full N-body
simulations.  The major shortcoming of this approach is that
representing the halo with a simple analytical potential is
unrealistic. Although recent studies have assumed slightly more
complicated forms for the potential, such as axisymmetric NFW profiles
\citep{Penarrubia:2006} or triaxial NFW profiles \citep{Law:2009},
they fail to include a realistic time evolution, as haloes grow in
stages through mergers, or to account for changes in triaxiality with
radius \citep{Hayashi:2007} and time.

In this paper, we present a more advanced approach for representing the potential of a halo using a series expansion. Our approach is based on the formulation of the self-consistent field (SCF) method \citep{Clutton-Brock:1973, Hernquist:1992}. The SCF method involves describing a density field as a series expansion and then using this to self-consistently evolve the field. This is usually done by representing the density field as an N-body particle sampling and integrating the orbits of the particles in the series expansion potential. Previous work has used this method to perform N-body simulations \citep{Weinberg:1996, Weinberg:1999} and recently it has been applied by \citet{Choi:2009} to simulate the potential of subhaloes. SCF codes (also known as expansion codes) have the advantage of being efficient, of scaling linearly with the number of particles and of suppressing small-scale noise. It is desirable that the lowest order radial basis function resembles the system of interest so that a large number of terms are not required just to describe the basic density distribution. This can be avoided by tailoring the basis functions to the system by numerically solving the Strum-Liouville equation for the particular density distribution (Weinberg 1999). We have not done this in this paper; instead, for simplicity we employ a radial basis function set based on the common simple analytical Hernquist halo profile \citep{Hernquist:1990}.

\par

Rather than using the SCF method for the purpose of performing a complete simulation, we use just the series expansion part of the technique to approximate a pre-computed evolving density field, in this case a dark matter halo. This halo expansion (HEX) method offers us the means to create realistic approximations of an existing time-varying halo, which can then be employed for resimulations. Our approach has the distinct advantage of providing a much more
realistic description of a halo potential than a simple static
analytical form, while still being inexpensive. The starting-point is a
full N-body simulation. A set of coefficients is calculated that
describes the halo with a chosen set of basis functions. Subsequently,
an estimate of the halo density or potential at any point in space can be obtained by evaluating the appropriately weighted sum of
the basis functions at that point. In addition, by calculating multiple independent sets of coefficients at various times in the halo's history and interpolating between the sets we can describe the halo at any time during this period.

\par

There is a wide range of possible applications of this method. It allows us to create approximations of very expensive halo simulations and then replay them at will. It can be used to study the evolving internal environment of the haloes or for the purpose of placing new objects into the simulations and observing behaviour as if they had been present in the original simulation. Problems to which it is ideally suited include: the orbits and
stripping of subhaloes, the response of a light disk to the changing
halo potential, the shape and precession of tidal streams, and the
dynamics of satellites galaxies. In this paper, we focus on the first
of these applications; we will explore the second in a later
paper. Comparing orbits within a halo approximated by a series expansion
to orbits calculated from the N-body halo serves as a demonstration of
the method and provides a test of the accuracy of the approximation.

\par

Limitations of our halo expansion technique include the lack of back reaction of the halo potential when new components are added. For example, if a model of a baryon disk is introduced, the associated reduction of the triaxiality at the centre of the dark halo \citep{Debattista:2008, Abadi:2010, Bett:2010} cannot be included in the expansion approximation. At present, the method does not treat the effect of dynamical friction on objects orbiting within the halo. Although this can, in principle, be
implemented in the method, \citet{Boylan-Kolchin:2008} find that, for
subhalo-to-halo mass ratios less than 0.1 the decay of the subhalo
orbit due to dynamical friction over a few Gyrs is small.

\par

This paper is organised as follows. Section~2 describes the theory
behind the expansion technique and how it has been applied to generate a
representation of the density and potential of a simulated dark matter
halo. Section~3 quantifies how well the approximation succeeds in
recreating the orbits of both single particles and subhaloes. The
latter part of the section carries out a comparison between the
evolution of subhaloes in a full simulation and in the approximated
potential. In Section~4, we use the expansion method for adding a new subhalo
into the halo and finally, in Section~5, we summarise our conclusions.

\section{Methodology}

We start by presenting a brief overview of the theory behind our expansion method based on the SCF formulation and then describe the simulated haloes to which it has been applied and the considerations required in its application.

\subsection{Basis Function Series Expansions}

The self-consistent field (SCF) method was originally devised by \citet{Ostriker:1968}, where it was used to find the equilibrium structure of rapidly rotating stars. \citet{Clutton-Brock:1972, Clutton-Brock:1973} applied the SCF method to computational stellar dynamics, to model the potential of simple galaxies. \citet{Hernquist:1992} (hereafter HO) further developed the technique and it is upon their formulation we base this paper. The idea of the SCF technique is to expand the density and potential in a set of basis functions. The coefficients for the density can be found by summing over the particle distribution of a simulation. The corresponding coefficients for the potential are then obtained through solving Poisson's equation. Differentiation of the potential series gives the acceleration, which can then be used to self-consistently evolve the particles. We adopt the SCF method for creating a series expansion for a N-body distribution but not use it to move the particles, instead we are interested in the expansion itself.

\par

We perform our expansion in spherical polar coordinates with $r$ the radial distance, $\theta$ the polar angle and $\phi$ the azimuthal angle. We start by considering the potential and density written as the biorthogonal series
\begin{align}
\label{eq:density_expansion_a}
\rho(r, \theta, \phi) & = \sum_{nlm} A_{nlm} \rho_{nlm}(r, \theta, \phi), \\
\label{eq:potential_expansion_a}
\Phi(r, \theta, \phi) & = \sum_{nlm} A_{nlm} \Phi_{nlm}(r, \theta, \phi),
\end{align}
where $ \rho_{nlm}(r, \theta, \phi)$ and $\Phi_{nlm}(r, \theta, \phi)$ are the basis functions labelled by $n,l,m$. A pair of biorthogonal series are defined by the property that
\begin{equation}
\int \rho(\vect{r})_{nlm} \Phi(\vect{r})_{n'l'm'} d\vect{r} = \delta_{nn'}  \delta_{ll'}  \delta_{mm'}.
\end{equation}
If the individual basis function series are not orthogonal then it is necessary to use a pair of biorthogonal series instead. When taking the overlap of the density with the potential basis functions, the biorthogonality property ensures that each coefficient only depends on a single potential basis function and that there is no contribution to it from any of the other basis functions. The basis functions are chosen so that each pair of terms are a solution to Poisson's equation
\begin{equation}
\mathbf{\nabla}^2 \Phi_{nlm}(r, \theta, \phi) = 4 \pi G \rho_{nlm}(r, \theta, \phi),
\end{equation}
with $G$ the universal gravitational constant.

\par

While we have a free choice of basis functions, it is desirable that lowest order terms be a good approximation to the system being modelled. This reduces the need to expand to high order to obtain a good fit. We have adopted basis functions from HO, where radial basis functions are based on the Hernquist profile. A Hernquist profile is a reasonable fit to a dark matter halo, having an appropriate slope of $r^{-1}$ at small radii but differing from the standard NFW form in its behaviour at large radii. For near spherical distributions it is natural to expand in spherical coordinates and use spherical harmonics. Equations (\ref{eq:density_expansion_a}) and (\ref{eq:potential_expansion_a}) then become
\begin{align}
\rho(r, \theta, \phi) & = \sum_{nlm} A_{nlm} \rho_{nl}(r) Y_{lm}(\theta, \phi),\\
\Phi(r, \theta, \phi) & = \sum_{nlm} A_{nlm} \Phi_{nl}(r) Y_{lm}(\theta, \phi),
\end{align}
where $Y_{lm}(\theta, \phi)$ are usual spherical harmonics. The zeroth order radial basis function is just the Hernquist profile
\begin{equation}
\rho_{00} = \frac{1}{2\pi} \frac{1}{r} \frac{1}{(1+r)^3},
\end{equation}
with potential
\begin{equation}
\Phi_{00} = - \frac{1}{1+r},
\end{equation}
when written in dimensionless units where $G=1$ and the scalelength in the Hernquist form, $a=1$. Higher order terms with $n=0$ result from the assumption that they behave asymptotically as $r\to\infty$ as would a usual multipole expansion. To construct terms with $n\neq0$ an additional radial function, $W_{nl}(\xi)$, is included, the form of which is found by it inserting into Poisson's equation. The transformation
\begin{equation}
\xi = \frac{r-1}{r+1},
\end{equation}
maps $r$ from the semi-infinite range to a finite interval and simplifies the following expressions.  Following the derivation from HO, the final full set of potential and density basis functions are finally found to be
\begin{equation}
\rho_{nl}(r) = \frac{K_{nl}}{2\pi} \frac{r^l}{r(1+r)^{2l+3}} C^{(2l+3/2)}_{n}(\xi) \sqrt{4\pi},
\end{equation}
and
\begin{equation}
\Phi_{nl}(r) = -\frac{r^l}{(1+r)^{2l+1}} C^{(2l+3/2)}_{n}(\xi) \sqrt{4\pi},
\end{equation}
where 
\begin{equation}
K_{nl} = \tfrac{1}{2}n(n + 4l + 3) + (l+1)(2l + 1),
\end{equation}
and $C^{(2l+3/2)}_{n}(\xi)$ are the ultraspherical polynomials \citep{Abramowitz:1964}. The expansions can then be rewritten in purely real quantities as 
\begin{align}
\label{eqn:density_expansion}
\rho(r, \theta, \phi) & = \sum^{\infty}_{l=0} \sum^l_{m=0}  \sum^{\infty}_{n=0} Y_{lm}( \theta) \rho_{nl}(r)[S_{nlm} \cos m\phi \nonumber \\ & + T_{nlm} \sin m\phi], \\
\label{eqn:potential_expansion}
\Phi(r, \theta, \phi) & = \sum^{\infty}_{l=0} \sum^l_{m=0}  \sum^{\infty}_{n=0} Y_{lm}( \theta) \Phi_{nl}(r)[S_{nlm} \cos m\phi \nonumber \\ & + T_{nlm} \sin m\phi].
\end{align}

\par

For a known density profile the expansion coefficients $S_{nlm}$ (or $T_{nlm}$) can easily be obtained by multiplying both sides of equation (\ref{eqn:density_expansion}) by $[Y_{lm}( \theta) \Phi_{nl}(r) \cos\phi]$ (or $[Y_{lm}( \theta) \Phi_{nl}(r) \sin\phi]$) and integrating over all space. This needs to be modified for N-body simulations where the density field is represented by discrete particles. In this case the integration over space becomes a sum over the particles, each weighted by its mass. Then the expansion coefficients are
 \begin{equation}
\left(\!\!
  \begin{array}{c}
    S_{nlm} \\ T_{nlm}
  \end{array}
  \!\!\right) = (2 - \delta_{m0}) \tilde{A}_{nl}\sum_k m_k \Phi_{nl}(r_k)Y_{lm}( \theta_k)
\left(\!\!
  \begin{array}{c}
   \cos m\phi_k \\\sin m\phi_k
  \end{array}
  \!\!\right),
\end{equation}
where
\begin{equation}
\tilde{A}_{nl} = -\frac{2^{8l+6}}{4 \pi K_{nl}}\frac{n!(n+2l+\tfrac{3}{2})[\Gamma(2l+\tfrac{3}{2})]^2}{\Gamma(n+4l+3)},
\end{equation}
and $r_k$ is the position of each particle and $m_k$ its mass.

\par
 
Once the coefficients are calculated, they can be used to evaluate equation (\ref{eqn:potential_expansion}) and find the potential at any location in space. Accelerations are obtained by differentiating the potential. By taking the gradient of equation (\ref{eqn:potential_expansion}) the accelerations can be written in spherical coordinates as 
\begin{align}
a_r(r, \theta, \phi) & = - \sum^{\infty}_{l=0} \sum^l_{m=0}  \sum^{\infty}_{n=0} Y_{lm}( \theta) \frac{d}{dr}{\Phi}_{nl}(r) \nonumber \\ & [S_{nlm} \cos m\phi + T_{nlm} \sin m\phi], \\
a_{\theta}(r, \theta, \phi) & = -\frac{1}{r} \sum^{\infty}_{l=0} \sum^l_{m=0}  \sum^{\infty}_{n=0} \frac{dY_{lm}( \theta)}{d\theta}{\Phi}_{nl}(r) \nonumber \\ & [S_{nlm} \cos m\phi + T_{nlm} \sin m\phi], \\
a_{\phi}(r, \theta, \phi) & = -\frac{1}{r} \sum^{\infty}_{l=0} \sum^l_{m=0}  \sum^{\infty}_{n=0} \frac{m Y_{lm}( \theta)}{\sin \theta}{\Phi}_{nl}(r) \nonumber \\ & [T_{nlm} \cos m\phi - S_{nlm} \sin m\phi].
\end{align}
Both the radial and spherical harmonic basis sets are complete, so when summed from $n=0 \rightarrow \infty$ and $l=0 \rightarrow \infty$ the expansion converges to the exact distribution, although non-uniformly near discontinuities. However, in practice the expansions are truncated at some high order term, $n_{\rm max}$ and $l_{\rm max}$.  Truncated to a finite number of terms, equations (\ref{eqn:density_expansion}) and (\ref{eqn:potential_expansion}) become
\begin{align}
\label{eqn:density_expansion_lim}
\rho(r, \theta, \phi) & = \sum^{n_{\rm max}}_{n=0} \sum^{l_{\rm max}}_{l=0}  \sum^l_{m=0} Y_{lm}( \theta){\rho}_{nl}(r)[S_{nlm} \cos m\phi \nonumber \\ & + T_{nlm} \sin m\phi],  \\
\label{eqn:potential_expansion_lim}
\Phi(r, \theta, \phi) & = \sum^{n_{\rm max}}_{n=0} \sum^{l_{\rm max}}_{l=0}  \sum^l_{m=0} Y_{lm}( \theta){\Phi}_{nl}(r)[S_{nlm} \cos m\phi \nonumber \\ & + T_{nlm} \sin m\phi],
\end{align}
with the number of terms determining the accuracy to which the expansions reproduce the actual density distribution.

\par

This algorithm is ideally suited to parallel computation. Each processor can independently calculate the coefficients for disjoint subsets of particles. A final summation collects together the contributions from each processor to generate the coefficients for the complete particle set. This ease of parallelism coupled with the algorithm being of $O(n)$ in the number of particles means it is ideally suited for use on huge datasets. However, the algorithm is to leading order $O(n_{\rm max}l_{\rm max}^2)$ for the number of basis terms included in the expansion and can quickly become computationally expensive if too many higher order terms are included.

\subsection{Simulations}

\begin{table*}
\center
\begin{tabular}{l l l l l l l l l l l}\hline \hline
Halo       &$m_{\rm p}$     &$\epsilon_{G}$&$r_{200}$&   $M_{200}$   &$N_{200}$&$V_{\rm max}$&$r_{\rm max}$ \\
           &  [$M_{\odot}$]  &  [pc$$]&[kpc$$]&[$M_{\odot}$]& [$10^6$]&  [km/s]        &  [kpc]      \\ \hline

Aq-A-2     &  1.370$\times 10^4$  &  66  & 244.84   & 1.842$\times 10^{12}$   & 134.47    & 208.49   &  28.14 \\
Aq-A-4     &  3.929$\times 10^5$  &  342 & 245.70   & 1.838$\times 10^{12}$   & 4.68      & 209.24   &  28.19  \\

\hline
\end{tabular}
\caption{Basic parameters of the two Aquarius simulations of the A
halo.  $m_{\rm p}$ is the particle mass in the high-resolution region,
$\epsilon_G$ is the Plummer-equivalent gravitational softening length,
$r_{200}$ is the  virial radius, defined as the radius enclosing a
mean overdensity 200 times the critical value, $M_{\rm 200}$ is the
mass within the virial radius, $N_{200}$ is the total number of
particles within $r_{200}$. Also listed is the position ($r_{\rm
max}$) of the peak ($V_{\rm max}$) of the circular velocity profile. 
  \label{tab:Aq}}
\end{table*}

This work is based on a simulated Milky Way sized dark matter halo from the Aquarius project \citep{Springel:2008a, Springel:2008b, Navarro:2010}. The Aquarius project sample consists of six haloes of mass  $\sim 10^{12}M_{\odot}$, which have each been resimulated at multiple resolutions. The simulations were performed using an improved version of {\sc gadget} \citep{Springel:2001b, Springel:2005}. The cosmological model used in the simulations assumes a $\Lambda$CDM cosmogony, with parameters: $\Omega_m = 0.25$, $\Omega_{\Lambda} = 0.75$, $\sigma_8 = 0.9$, $n_s = 1$ and Hubble constant $H_0 = 73$ kms$^{-1}$ Mpc$^{-1}$. The six haloes were selected from the set of all isolated $\sim 10^{12}M_{\odot}$ haloes from a lower resolution $900^3$-particle parent simulation of a $100 h^{-1}$ Mpc box. Isolated means that a halo had no neighbours exceeding half its mass within $1 h^{-1}$ Mpc; this ensured that the haloes were not members of any massive groups or clusters. Gravitationally bound substructures orbiting within the main larger Aquarius haloes are identified using the SUBFIND algorithm \citep{Springel:2001a}. 

\par

The Aquarius project haloes are ideally suited for this work as they are high-resolution simulations of single haloes, that have been carefully tested for convergence and have a large number of outputs saved at regular times. We have applied the expansion technique to two different resolution versions of the Aquarius A halo. The majority of this work is based on the higher resolution version known as Aq-A-2, while a lower resolution version, Aq-A-4, is used to check for convergence. Table \ref{tab:Aq} details the basic parameters of the simulations and haloes. There is a factor of 28 difference in the resolution of the two versions, with excellent convergence found between them. The Aq-A-2 simulation has a total of 1024 outputs, while the Aq-A-4 has only 128. For this work we have restricted ourselves to the same 128 outputs from both versions, giving one approximately every 155 Myrs at late times.

\subsection{Application to Simulated Haloes}

To apply the HEX technique to a dark matter halo from the Aquarius simulation, we expand about the potential minimum, as identified by SUBFIND by the most bound particle. A summation over all particles is performed, once for each halo, to yield a set of coefficients that describe the halo by the given basis functions. We limit the expansions to a small number of terms, resulting in a set of coefficients much smaller in comparison to the number of dark matter particles in the halo. This truncation of the series smooths the density and removes small-scale detail. 

\par

Only particles within 1.3 virial radii of the halo centre are included in the coefficient summation. At greater distances, the distribution of material is more irregular and not well fitted by spherical basis functions. While the use of a hard cut-off at the boundary imposes a discontinuity in the density profile there, we find this not to be a problem. We have tested with larger, as well as soft boundaries and find the exact choice makes little difference to our results. We choose to use a hard boundary at for simplicity.

\begin{figure}
\centerline{\includegraphics[width=1.2\linewidth]{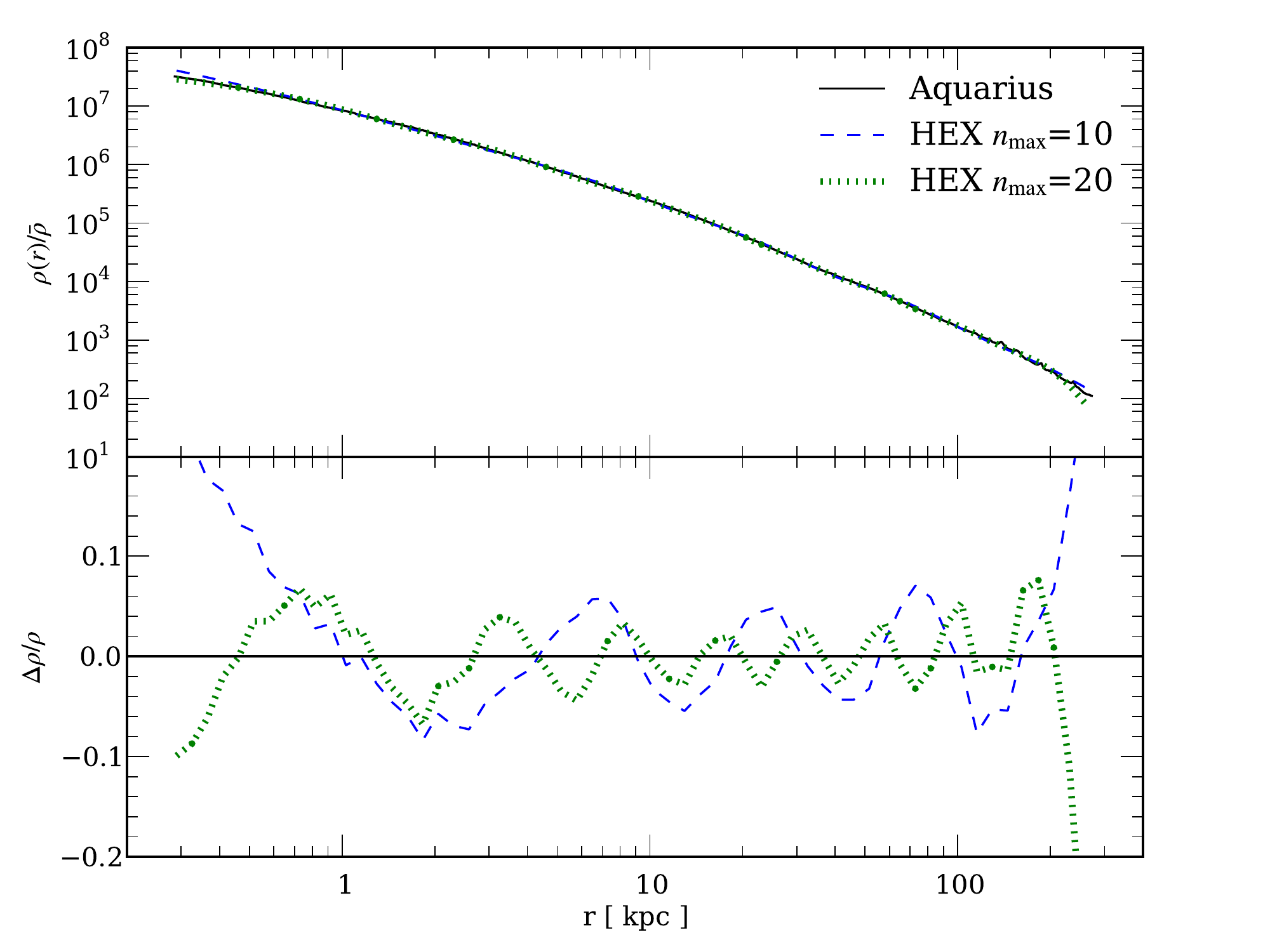}}
\caption{\label{fig:densityProfile}\textit{Upper panel:} Spherically-averaged density profiles $\rho(r)$ of the main Aq-A-2 halo. The solid line is the profile of the actual halo from the simulation, while the dotted and dashed lines are the profiles from the expansion with $n_{\rm max}=10$ and $n_{\rm max}=20$ respectively. \textit{Bottom panel:} Residuals of the density profile fits, $\Delta \rho/ \rho \equiv (\rho_{\rm HEX} - \rho_{\rm halo}) / \rho_{\rm halo}$, where $\rho_{\rm halo}$ is the true halo density and $\rho_{\rm HEX}$ denotes the HEX approximated density.}
\end{figure}

Fig. \ref{fig:densityProfile} shows the comparison of the density profile of the main halo from the Aq-A-2 simulation, obtained by binning the simulation particles into spherical shells, with its approximation by the HEX method. The lower panel shows the residuals between the model and the data. It can be seen that over the radial range 1-100 kpc, using just eleven radial basis functions, $n_{\rm max}=10$, the RMS deviation of the residuals is $4.2\%$, decreasing to $2.6\%$ when twice the number of radial terms, $n_{\rm max}=20$, are included. Even using just a few radial basis functions the expansion achieves a fit to within a few percent to the spherically averaged density profile of the halo, over a range where the radial density varies by over six orders of magnitude.

\subsubsection{Order of Expansion}

\begin{figure}
\centerline{\includegraphics[width=1.2\linewidth]{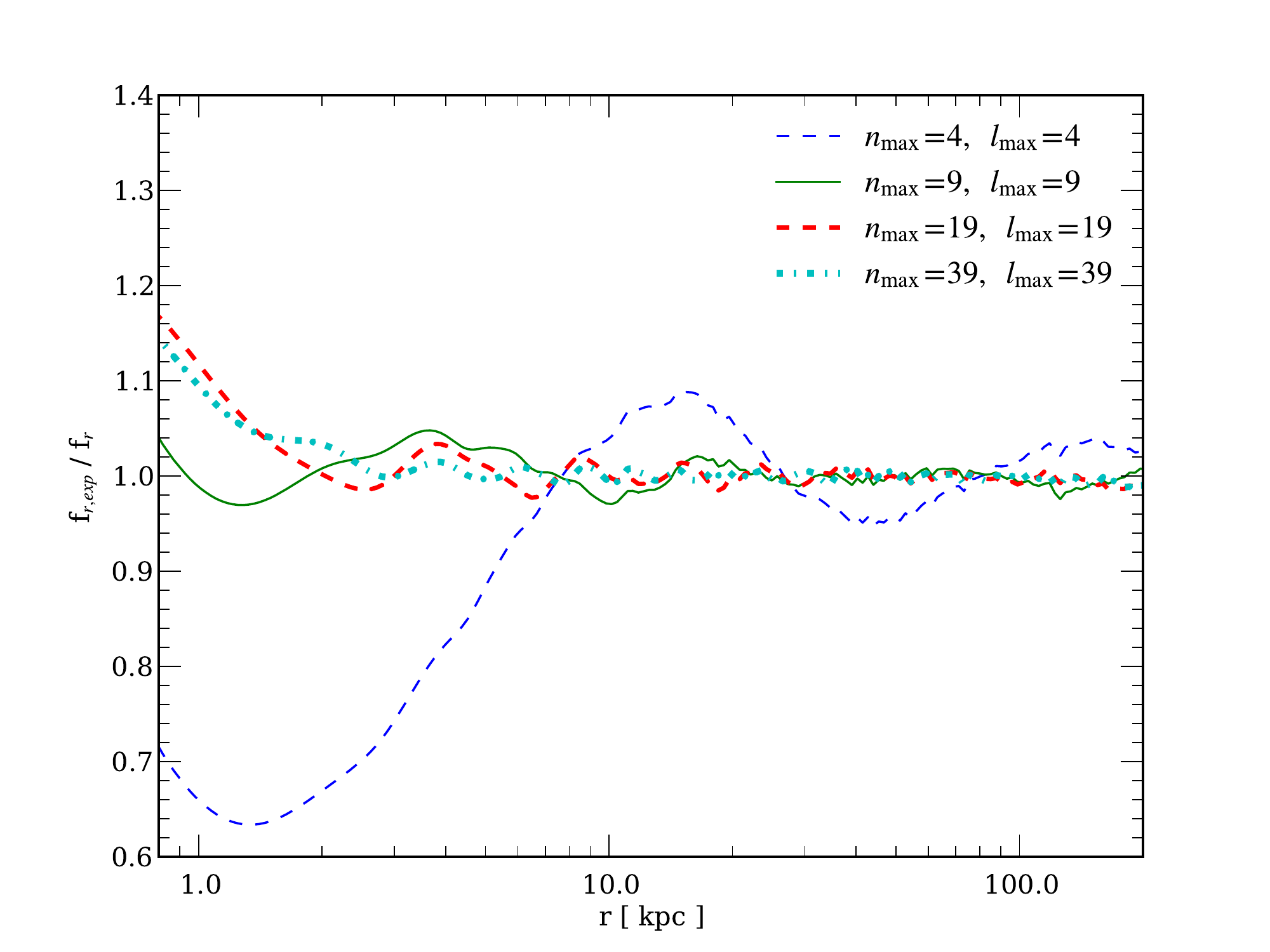}}
\caption{\label{fig:forceComparison} Radial component of the force calculated from the HEX approximation truncated at differing $n_{\rm max}$ divided by the actual force calculated directly from the Aq-A-4 simulation.}
\end{figure}

The accuracy of the approximation of the halo depends on the number of terms included in the expansion, the use of more terms allows smaller spatial features to be resolved. The spatial resolution approximately scales inversely proportional to $n_{\rm max}$ and $l_{\rm max}^2$. The effect on the force of including more terms in the expansion can be seen in Fig. \ref{fig:forceComparison}. Here, the radial component of the force for $n_{\rm max} =  l_{\rm max} = 4, 9, 19, 39$ is compared to the force as calculated directly from the original N-body simulation. 

\par

In the central region of the haloes the radial force estimated from the expansion differs from that calculated in the simulations. The closer to the centre, the larger the disagreement. This divergence is due to the density of the simulated halo having a logarithmic slope shallower than -1, while the lowest order Hernquist basis function having a cusp at the centre with a slope of -1 and not being a good fit there. Excluding the centre from the comparison , so considering the region between 5 and 100 kpc, it is found that doubling both $n_{\rm max}$ and $l_{\rm max}$ from five to ten terms results in a big improvement, with the fractional RMS deviation falling from $4.8\%$ to $1.3\%$. Doubling the number of terms again gives further gains, with expansions using 20 and 40 terms resulting in fractional RMS deviations of $0.83\%$ and $0.46\%$ respectively.

\par

As the expansion is taken to increasingly higher orders, the contribution of individual terms declines. Higher order terms resolve smaller scale structure, and eventually the very high order terms model only the shot noise from the discrete particle nature of the simulation. Following \citet{Weinberg:1996}, we take the signal-to-noise on a coefficient as $S/N\equiv [S_{nlm}^2/{\rm var}(S_{nlm})]^{1/2}$ , where by considering the computation of the coefficients as a Monte Carlo integration the variance can be estimated. Signal-to-noise of less than one indicates that the particle distribution does not provide significant information on the value of that coefficient. We find that terms even as high order as $n_{\rm max} = l_{\rm max} = 20$ enjoy low levels of noise and contribute to resolving the halo structure. This is not surprising as the Aq-A-2 has over 100 million particles within the virial radius, while an expansion with $n_{\rm max} = l_{\rm max} = 20$ only contains 8000 terms.

\par

Gravity is a long-range force dominated by the large-scale distribution of material. The force on an object is therefore determined primarily by the overall distribution of mass, and resolving nearby small-scale fluctuations does not substantially improve the radial force estimate. Going to higher expansion orders is thus unnecessary, as long as we employ sufficient terms to resolve the large-scale structure. Additional terms do not provide much gain. A force accuracy of less than $1\%$ can be achieved using $n_{\rm max} = l_{\rm max} = 20$, and is sufficient for most purposes. We use expansions to this order in the rest of this paper.

\subsubsection{Choosing the Scalelength}

The adopted set of basis functions contains a single free parameter corresponding to the scalelength, $a$, of their underlying Hernquist profile. This scalelength needs to be predetermined and chosen so that the lowest order basis function is a good fit to the halo. We find that the accuracy of the expansion when approximating the force is fairly insensitive to the exact choice of scalelength. Examination of the RMS deviation in the radial force as a function of scalelength shows that very small scalelengths give bad fits to the profile but any scalelength larger than 10 kpc is acceptable, with a minimum RMS deviation at 33 kpc. As we have already seen, the lowest order basis function is not a good fit to the halo at the origin due to a mismatch in central slopes. Reducing the scalelength does not help this.

\par

The basis functions are primarily constrained by the region where the number of particles per radial interval is a maximum. This occurs where the logarithmic slope of the density profile is -2, which is at the scalelength for an NFW profile and at half the scalelength for a Hernquist profile. It is in this region that we desire the lowest order basis functions to fit well in order to minimise the number of terms needed in the expansion. The Aquarius A halo is very well fitted at $z=0$ by an NFW profile with a scalelength of 15.3 kpc. It is therefore unsurprising that the optimum scalelength for the best fit by the lowest order Hernquist basis function is found to be 33 kpc, approximately twice the the best fit NFW scalelength. Using this value obtains an average RMS deviation in the radial force between 5 and 100 kpc of 0.53\%, with the force correct to within $3\%$ down to 2 kpc. In the rest of the paper we use a scalelength of 33 kpc when modelling this halo.

\subsubsection{Frame of Reference}

We perform the expansion in a frame moving with the halo. Haloes are accelerated by surrounding large-scale structure. In the simulation this results in the halo having a peculiar velocity of several hundred kilometres per second, a velocity comparable to the relative motion of material within it. We wish to transform into a frame in which we can treat the halo as stationary. This will allow us to follow the relative motion of objects within a halo, such as the orbit of particles, and neglect the halo's movement through space in their equation of motion and not need to take into account the position or the velocity of the halo at intermediate times. Because of the halo's finite extent, this frame is not strictly an inertial frame, but is accelerating due to the gravitational effects of distant large-scale material. Since material within the halo experiences the same acceleration, this is only important if there are significant differential tidal forces over the scale of the halo, but this is not the case; the long-range tidal force, calculated by direct summation, from distant material is less than $1\%$ the magnitude of the internal halo force within 100 kpc of the halo centre and can be safely ignored. 
 
\par

In order to transform into a stationary halo frame we must define an origin that moves with the halo and remove the halo velocity. The origin of the halo frame is chosen as the halo potential minimum, $\vect{x}_{pm}$. This is a well defined point that follows a smooth path. The choice of the halo velocity to use for the transformation to a stationary frame is not obvious. We need to use the instantaneous halo velocity to make the correct transformation rather than the average velocity, which we could simply obtain from the motion of the potential minimum. A sensible choice is to look at the net motion of the material surrounding the potential minimum. We obtain a centre of mass velocity that corresponds to the potential minimum's velocity by selecting all particles within some bounding radius, $R$, of the halo centre. The velocity is then
\begin{equation}
\vect{v}_c = \frac{\sum_i m_i \vect{v}_i}{\sum_i m_i}
\end{equation}
where $i$ is all the particles that have $\lvert \vect{x}_i-\vect{x}_{pm} \rvert \le R$. Restricting ourselves to the just inner region where the halo is almost static we find that the exact choice of $R$ makes little difference to the centre of mass velocity. Varying $R$ between 1 and 20 kpc alters the velocity by less than a kilometre per second. Including the entire halo gives a centre of mass velocity some 20 kms$^{-1}$ different from that of the inner regions. We therefore choose to use the centre of mass velocity of the particles within 5\% of the virial radius, which for Aq-A-2 is $R = 12$ kpc at $z = 0$.

\par

To show that this is a valid choice we compare the orbits of particles integrated within the expansion to the orbits the same particles took within the original simulation. The next section describes this in detail. We find that for each subset of particles there is an optimal choice of velocity for the halo frame in which to integrate particle orbits in order to match their equivalent orbits from the Aquarius simulation. This velocity can be found through a minimisation scheme, in which we attempt to minimise the difference in their final position compared to their position in the original simulation. While the optimum velocity depends on the set of particles considered, it only varies within a few kilometres per second between cases, suggesting that the motion of the inner regions of the halo is almost uniform. A slight difference in motion throughout the halo is the cause of the small spread and allows us to define a window of several kilometres per second in which we find that any choice of velocity for the halo frame works satisfactorily. Choosing a different velocity within this window changes the path of the orbital integration by only a percent or two. 

\par

Not only does this show that an approximately stationary frame does exist, we also find that the centre of mass velocity that we chose earlier lies within this window. This is true for the Aquarius A, B and C haloes and demonstrates this to be a valid choice for the halo frame, especially as it can be easily determined in advance, whereas the optimum velocity for a particular case can only be located retrospectively. The resulting procedure for placing objects within the expansion approximation frame is to find their initial position relative to the halo potential minimum at the start time and their initial velocity with respect to the defined halo velocity, $\vect{v}_c$. The motion of the objects can then be followed totally within this frame and there is no need to further consider the overall motion of the halo.

\subsubsection{Time Variation}

\begin{figure}
\centerline{\includegraphics[width=1.2\linewidth]{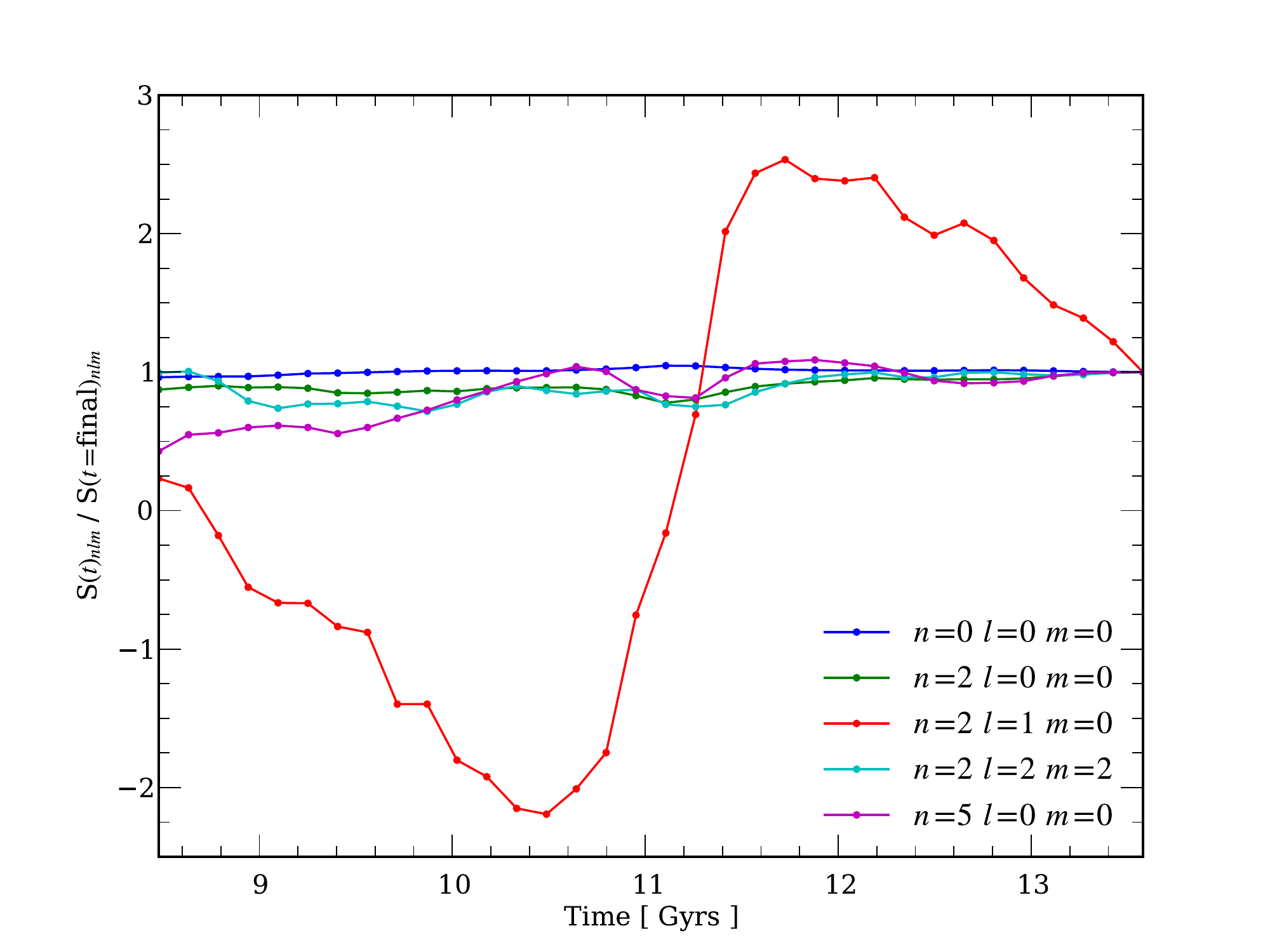}}
\caption{\label{fig:timeVariation} The variation of low order coefficients as a function of time for the last 5 Gyrs of the Aq-A-2 halo evolution.}
\end{figure}

Due to the fact that such a large amount of data is generated, the output of N-body simulations is usually recorded only at a few snapshots. Between these snapshots information on the exact evolution of the halo is lost. However, it is usually sufficient to use simple interpolation to approximate it. The expansion technique is ideal for this because at each snapshot a new set of coefficients are calculated to describe the halo at that time. An approximation to the state of the halo at any intermediate time can be recovered by linearly interpolating the coefficient of each basis function between the directly preceding and following snapshots.

\par

Fig. \ref{fig:timeVariation} illustrates the variation in a selection of low order coefficients over the last 5 Gyrs of the Aq-A-2 halo's growth, with a time resolution set by the snapshot spacing, of 155 Myrs. The coefficient of the lowest order basis function varies very little, initially showing a slight increase until 11 Gyrs, followed by a slight decline. The variation corresponds to the slight fluctuation in mass of the inner $\sim$100 kpc of the halo. The higher order coefficients have greater variation. The fluctuations on short timescales, of the order of the time spacing of the snapshots, are generally small, while the larger, more important, variations, such as the oscillation in the $n=2$, $l=1$, $m=0$ coefficient, occur on longer timescales. The time spacing we use is sufficient to capture large-scale changes in halo structure. Smaller quicker changes, such as those from substructure, may be missed but this does not matter as these are not spatially resolved by the expansion anyway.

\section{Resimulating Aquarius}

Once we have obtained a time-varying set of coefficients for a halo expansion approximation of an Aquarius halo potential and density, it is straightforward to use this to integrate orbits of test particles within the evolving halo potential. In order to test the accuracy of the HEX method, we examine how closely we can reproduce the properties of existing objects already present in the Aquarius simulations along their orbits. Based on the findings of the previous section we use a potential expansion including terms up to order $n_{\rm max} = 20$ and $l_{\rm max} = 20$, with a fixed scalelength of 33 kpc and summed over all particles within 340 kpc of the halo centre, to approximate Aq-A-2 halo. A set of coefficients is generated for each snapshot, approximately every 155 Myrs. 

\subsection{Integrating Orbits}

Ideally, if the potential is approximated accurately, test particles placed in the evolving halo potential will behave in the same manner as particles in the original simulation. This should be the case as long as the particles are not bound to any subhalo, since we are not attempting to resolve this level of detail. Therefore, by setting up a test particle with initial conditions matching the instantaneous state of a simulation particle and integrating the orbit within the HEX approximation, a comparison can be made between the path that the simulation particle actually followed and the one recreated using the HEX method. Differences in the orbital path or properties provide a guide to the accuracy of the HEX approximation.

\begin{figure}
\centerline{\includegraphics[width=1.1\linewidth]{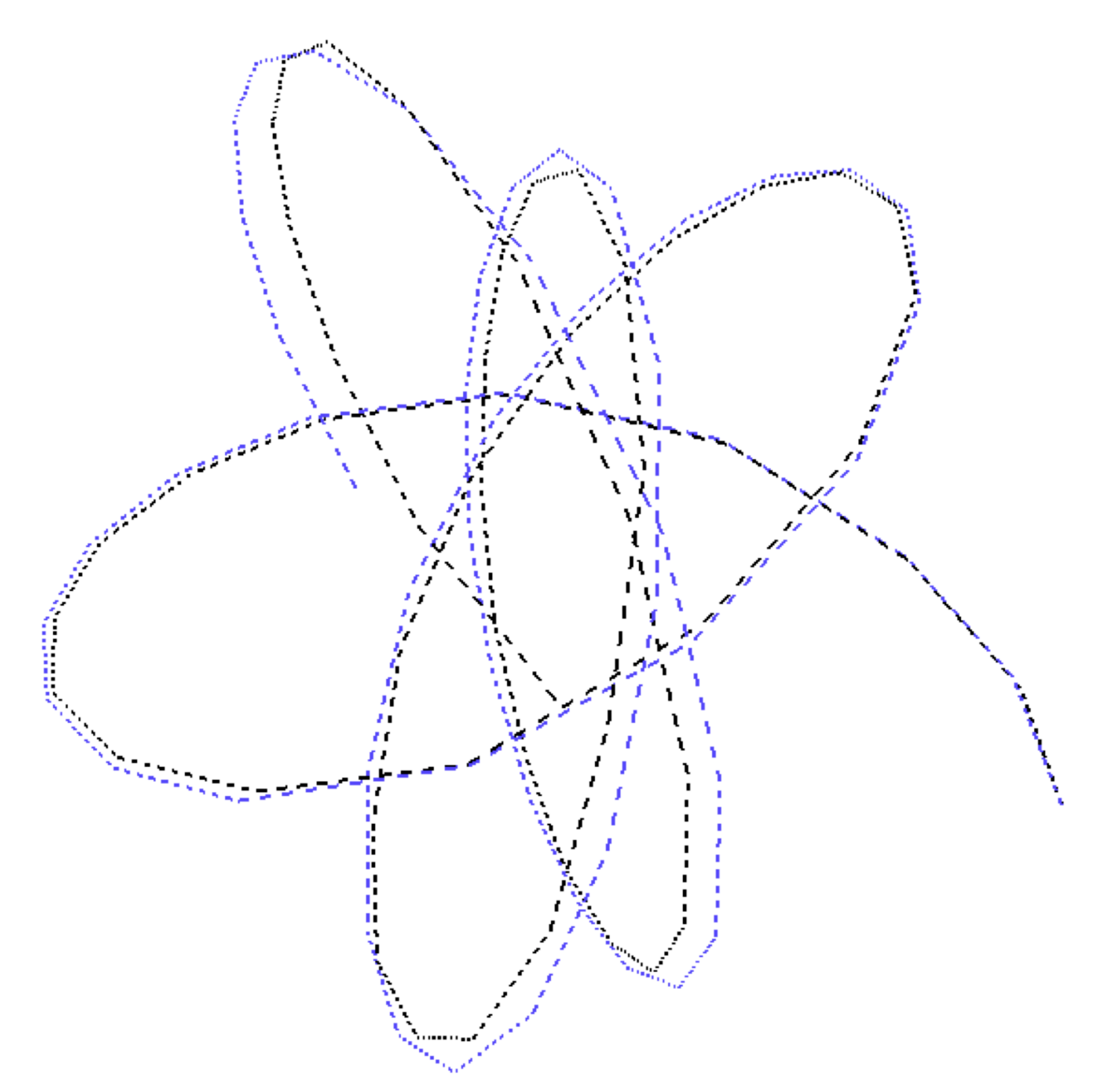}}
\caption{\label{fig:singleParticleOrbit} The orbit of a single particle in the Aq-A-4 simulation. The blue line shows the result of using the HEX approximation. The black line shows the actual path of the same particle followed through the Aquarius simulation. The particle positions were recorded only at limited points which have been joined by straight lines. Both paths start from the same point on the right and are integrated for 1 Gyr.}
\end{figure}

Fig. \ref{fig:singleParticleOrbit} shows an example of an orbit that is particularly well reproduced. The orbit of the particle extracted from the Aq-A-4 simulation is compared with one integrated for 1 Gyr in the HEX potential. The recreated orbit closely matches the actual particle path, though it slightly diverges over time. By the end of the integration there is some displacement between the final positions. While the orbit shape is well reproduced, the progress of the particles along their orbits is slightly out of phase. This discrepancy was introduced during the 3rd apocentre passage, when the resimulated particle took a slightly wider orbit so that it  subsequently lags behind the actual particle. An increasing divergence in paths is not unexpected as once a particle has even slightly different phase space coordinates it will subsequently follow an increasingly different orbit. The energy of the two particles is matched to within $1.3\%$ throughout the entire orbit.

\par

Once paths start to diverge, the particles will travel through different parts of the halo and it is therefore unsurprising that the properties of the original and recreated orbits become increasingly uncorrelated. It is more interesting to consider the properties of the particles over short time periods while the paths are still very similar. We do this for a set of 100 particles, randomly selected from the Aq-A-2 simulation from within 140 kpc of the halo centre at a redshift $z=0.5$.  We extract their orbits over 5 Gyrs by finding their positions through 33 successive snapshots. 

\par

In order to compare the acceleration of these particles in the HEX approximation to the acceleration they experienced in the original {\sc gadget} simulation, we must remove the overall halo acceleration from the {\sc gadget} values. This is necessary as the integration in the HEX approximation is performed in the moving halo frame. The linear component of the overall halo acceleration is easy to remove and shows up as a systematic offset in the accelerations between the two cases. Calculating the mean acceleration difference in the final $z=0$ snapshot finds a clear offset of 18.2 kms$^{-1}$ Gyr$^{-1}$. Once this component is removed we find a close match in the accelerations, with a median acceleration difference of 1.2\% for the 100 particles over 33 snapshots. 

\par

A comparison between the HEX approximation and a direct N-body force summation of the same material included in the HEX expansion gives a slightly better agreement for the median force differences of 0.9\%. The differences between this N-body summation and the {\sc gadget} force arise from a combination of the higher order acceleration components not being removed, possible errors in the force calculated by {\sc gadget} which come from a TreePM method, also an approximation, and the fact that the box containing the simulation is treated as periodic by {\sc gadget}. Regardless of these slight differences, both the comparison with {\sc gadget} calculated force and the direct summation demonstrate that there is in general an average force/acceleration error of approximately 1\% for the HEX approximation. In certain situations there can be much larger errors; in one case we find a difference of 90\%, when a particle comes within 500 pc of a large subhalo. Differences of this size are expected for the HEX potential near subhaloes, since such subhaloes are not well resolved in the approximation.

\par

Integrating the orbits of the test particle set over the short time period between snapshots allows us to measure the distance between the final positions and the actual particle positions in the Aquarius simulation. The integration is done by treating the particles as non-interacting and placing them at the same initial position and with the appropriate relative velocity, and using a simple drift-kick-drift leapfrog integrator with a fixed time step of 1 Myr. By using the difference in forces at the snapshot times as an estimate for the average force error we are able to calculate the expected divergence of orbits between snapshots and compare this to the divergence obtained from the HEX integration. Over the short time scale between snapshots of $\sim$ 155 Myrs the displacements are small, usually a few hundred pcs. We find that the error in the displacement of the integrated paths are consistent with the estimated error.

\subsubsection{Energy Changes}

Examining how well the HEX approximation reproduces the integrals of the motion can be more indicative of differences in orbits than looking at the differences in final position. Position is an instantaneous phase space coordinate that rapidly varies along an orbit, and absolute differences in position are dependent on a particle's current radial distance from the halo centre. In contrast, energy, although not an integral of the motion since the potential is time-varying, changes slowly along the orbit. In a spherical potential, angular momentum would also be an integral of the motion. However, the Aquarius haloes are strongly triaxial, particularly in the centre, so contain a significant number of box type orbits \citep{Binney:1987}. For these orbits the angular momentum varies rapidly over very short time-scales, which makes a comparison between the Aquarius simulation and the orbits integrated in the HEX approximation less useful. In this section we therefore only consider energy.

\par

\begin{figure}
\centerline{\includegraphics[width=1.1\linewidth]{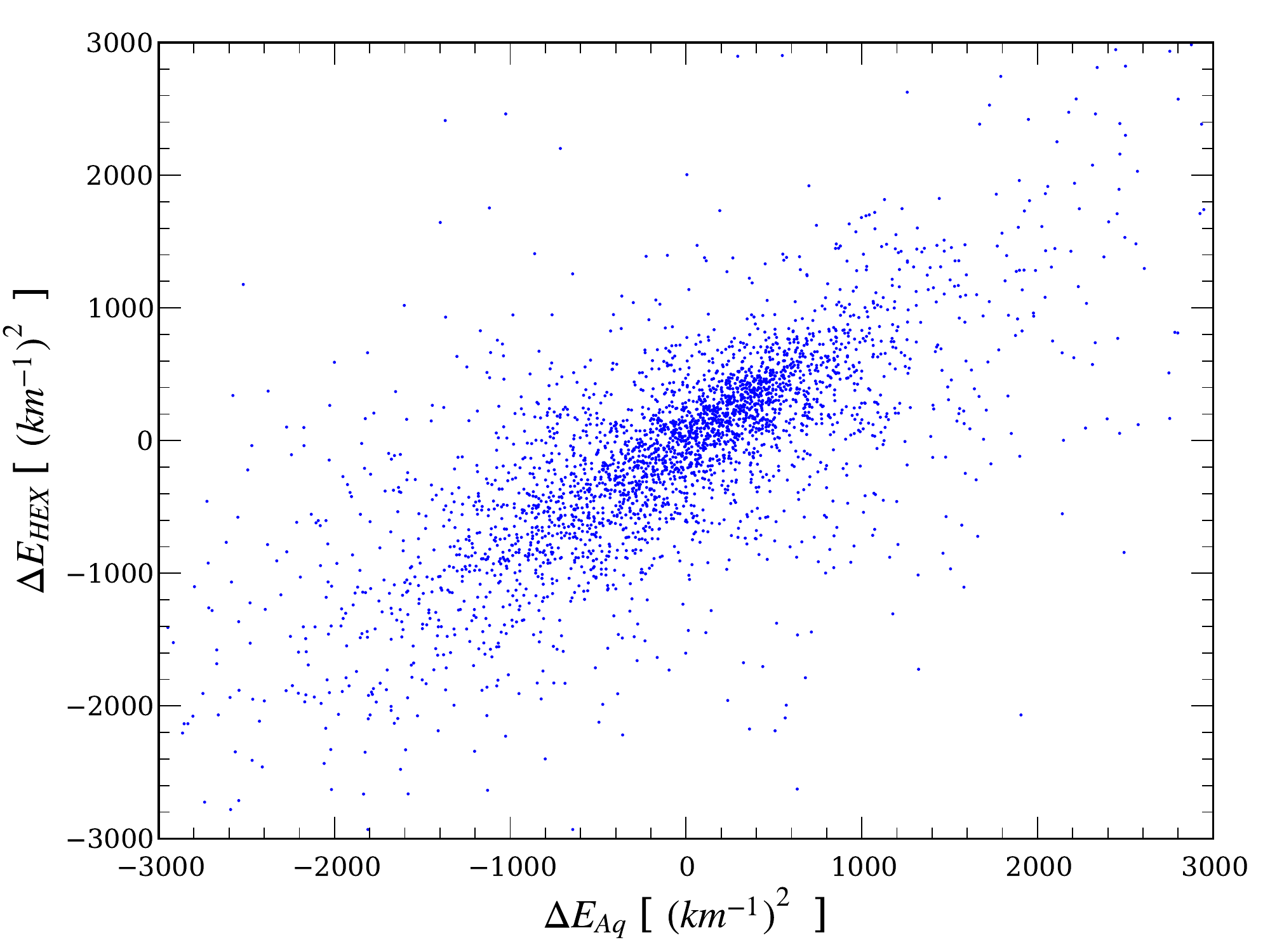}}
\caption{\label{fig:energyChange} Energy change for selected particles between snapshots in the Aquarius simulation compared to energy change for the same particles when their orbits are integrated over the same period in an HEX approximation of the Aq-A-2 halo.}
\end{figure}

By again integrating the particle orbits between snapshot times in the HEX approximation we can obtain a change in the orbital energy for each particle. We have checked that the change in energy is equal to the integral of the time variation of the potential along the path to within 1\%. Using a smaller step size has a negligible effect on our results, confirming that the changes in energy are not due to the numerical integration. Fig. \ref{fig:energyChange} shows the correlation between the changes in the particles' energy in the Aquarius simulation compared to their energy change in the integrated HEX potential. There is a clear correlation between the two cases, with a Pearson correlation coefficient of 0.75. The HEX approximation does well at reproducing energy changes even though the particles may not follow exactly the same paths. As well as path differences the linear interpolation between coefficients will give a different variation in the potential at intermediate times, however, this does not seem to be important.

\subsubsection{Encounters}

Some of the test particles' orbits are significantly different to their Aquarius counterparts; they initially follow the Aquarius orbits but suddenly diverge and take very different paths. This occurs primarily for particles with low angular momentum on nearly radial orbits. The pericentric passages of these orbits are very close to the centre of the halo. As the particles approach the centre, the separation distance between the reconstructed orbits and the Aquarius paths becomes of the same scale as the pericentric distance. The large relative path separation results in the paths having substantially different approach angles and substantially different impact parameters, even in some cases passing opposites sides of the centre. Since the centre is very strongly triaxial, the change in the angular momentum during the encounter with the non-spherical centre is sensitive to the direction of the incoming objects and will cause the pairs of particles to be diverted in radically different directions.

\par

As well as the centre, which is responsible for the majority of these divergences, encounters with subhaloes can have a similar effect. Particles can either be deflected by subhaloes or become bound to them. To properly resolve a subhalo 1 kpc in size and 50 kpc from the centre would require at least an expansion with $n_{\rm max} =  150$ and $l_{\rm max} = 150$, over 3 million terms. 

\subsubsection{Population Distribution}

\begin{figure}
\centerline{\includegraphics[width=1.1\linewidth]{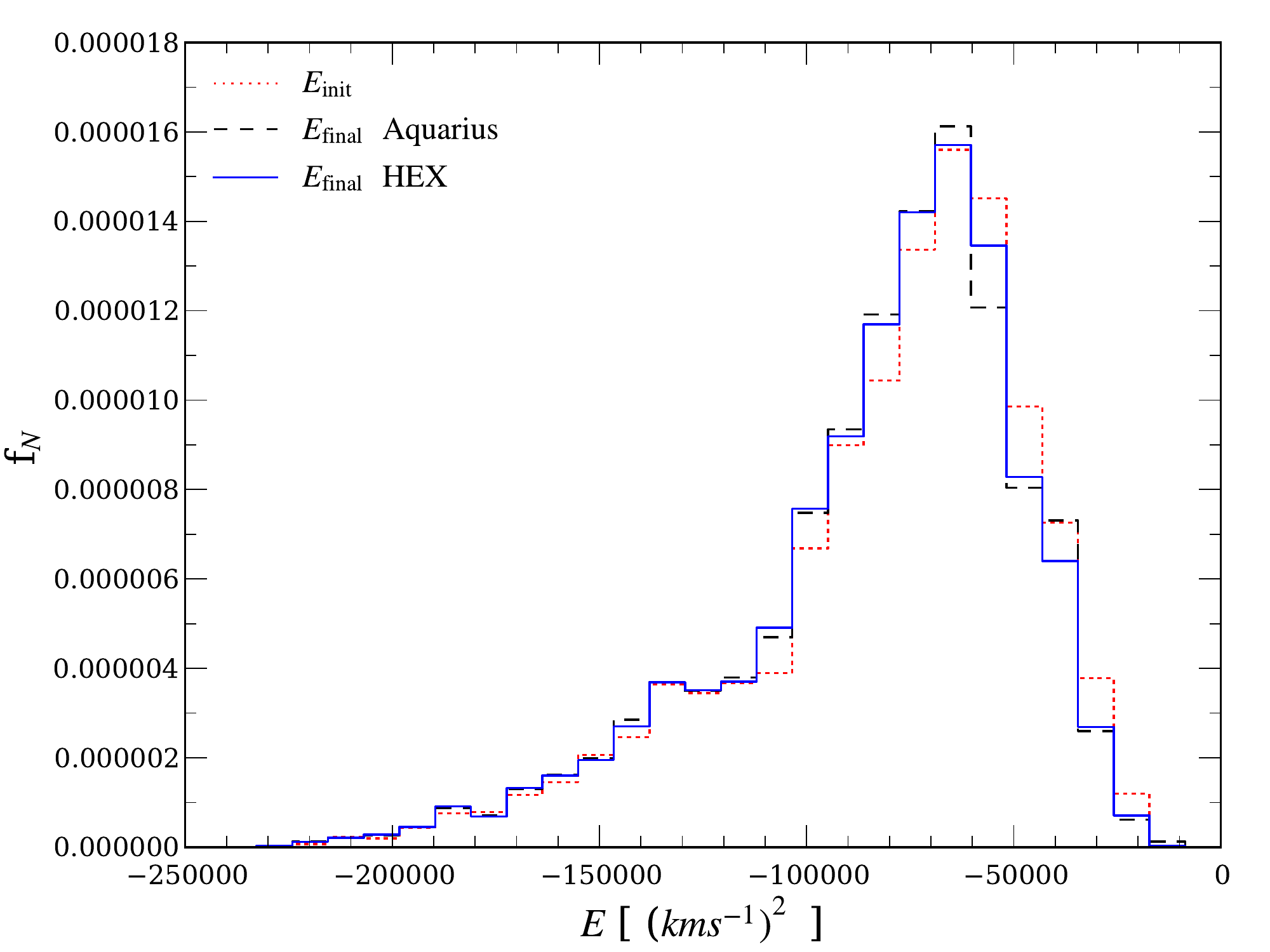}}
\caption{\label{fig:particlePopulationEn} The distribution of the total energies of the 10,000 test particles. The dotted line shows the initial energy distribution, while the dashed line is the distribution of their energies in the simulations after 5 Gyrs. The solid line is the energy distribution in the HEX resimulation.}
\end{figure}

\begin{figure}
\centerline{\includegraphics[width=1.1\linewidth]{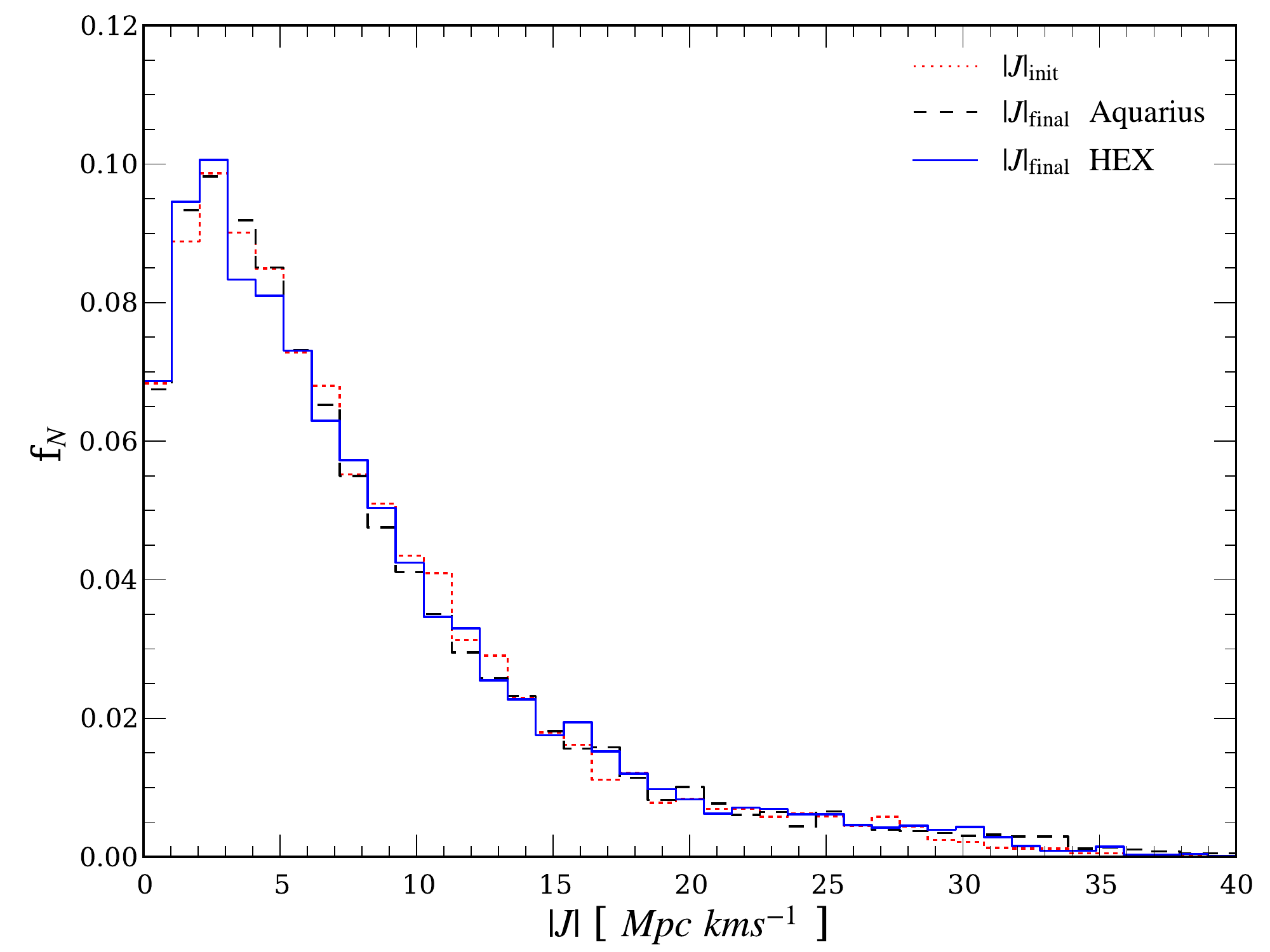}}
\caption{\label{fig:particlePopulationAM} The distribution of the magnitude of the angular momentum of the 10,000 test particles. The dotted line shows the initial distribution, while the dashed line is the distribution in the Aquarius simulations after 5 Gyrs. The solid line is the distribution in the HEX resimulation.}
\end{figure}

Even though individual orbits integrated in the HEX approximation may not always match their counterparts, the overall distributions of the energy and the magnitude of the angular momentum are well reproduced. This can be seen in Fig. \ref{fig:particlePopulationEn}, the distributions of total energies of 10,000 randomly selected test particles and Fig. \ref{fig:particlePopulationAM}, the distribution of the magnitude of the orbital angular moment. Both figures include the initial distributions and the final distributions from both the original Aquarius simulation and HEX resimulation over 5 Gyrs.

\par

The final energy distributions are very similar. A Kolmogorov-Smirnov test gives a probability of 0.24 that the energy distributions are drawn from the same parent distribution. Therefore, the null hypothesis that the energy distributions of the orbits from the Aquarius simulation and HEX resimulation are the same is not rejected at a statistically significant level. There is equally good agreement for the angular momentum, with a 0.42 K-S test probability. The very similar distributions suggest that while individual orbits may not be exactly reproduced, there is no systematic difference in orbits integrated in the HEX approximation and those found in the Aquarius simulation. There is, however, a significant difference between the final and initial distributions, with a K-S test probability of less than $1.3 \times 10^{-12}$ that the samples of orbital energies are drawn from the same distribution. The halo is accreting new material and evolving over the period of consideration, changing the overall distributions of energy. The fact that we match the final simulation distribution using the HEX approximation clearly demonstrates that the method correctly reproduces this evolution.

\par

Focusing on orbits confined near to the centre of the parent halo we find an even better match than ones with larger apocentric distances. This is a consequence of both the fact that the basis functions used in the HEX approximation have lower spatial resolution at larger radii and thus structure is not resolved as clearly in the outer regions, and the fact that the halo is dynamically older and more stable towards the centre. Restricting our attention to particles confined to a region near the centre of between 3 kpc and 20 kpc, where the HEX expansion is very successful, selects particles on near circular orbits. When we consider the energy and angular momentum distributions for these orbits, we find that there is little evolution in the distributions, with significant K-S probability of 0.14 for energy and 0.76 for angular momentum, that the population properties of the initial and final simulations have not changed. There is also very good agreement between the HEX and the simulation distributions, 0.97 for energy and 0.37 for angular momentum. Orbits in this region are of particular interest when considering galactic disks.

\subsection{Subhaloes}

Having studied the orbits of individual particles, we now turn our attention to the dynamics and evolution of subhaloes. These are large, gravitationally bound, extended bodies undergoing tidal evolution as they orbit within the parent halo. We compare the orbits of subhaloes resimulated within different halo expansion potentials, treating the subhaloes as extended objects, to the orbits of subhaloes from the Aquarius simulation. To model a subhalo as an extended body, we select the subhalo from the Aquarius simulation and identify all the particles that SUBFIND assigns to it. The same particles are extracted from subsequent snapshots and SUBFIND is run on just this particle set to calculate those that are still gravitationally bound. This results in a complete orbital path and record of the subhalo's evolutionary history. The resimulation of the subhaloes is done using a version of {\sc gadget} modified to allow additional HEX external potentials. The subhaloes are composed of multiple particles allowed to interact gravitationally. From the Aq-A-2 simulation we selected all 1507 subhaloes with 100 or more particles that are within 90 kpc of the centre of the parent halo at $z=0.5$. Their orbits and evolution are then integrated for 5 Gyrs in the HEX potential.

\par

The contribution to the potential from a subhalo needs to be removed from the halo expansion that is used to resimulate its orbit. Not excluding the self-contribution would lead to a double counting of the subhalo, because the gravitational effects of the subhalo are already included in the potential expansion. The double counting would generate an unrealistic self-attraction to the resimulated counterpart. Since the coefficients are just linear sums it is easy to remove the contribution from the subhalo by separately calculating the coefficients of just the subhalo particles from the original simulation and subtracting them from the total coefficients. This does not remove the entire presence of the subhalo from the HEX approximation, as the halo response (i.e the dynamical friction wake) is still part of the expansion. While a resimulated subhalo closely follows the same orbit as in the original simulation the wake can be an additional source of drag. However, an estimate of the dynamical friction on a subhalo based on the Chandrasekhar model \citep{Chandrasekhar:1943} shows that it is negligible for the majority of subhaloes and only really important for the most massive ones. 

\par

In contrast there can be no new halo response to the subhaloes in the resimulation, due to the fixed nature of the expansion, therefore there is no direct dynamical friction on the subhaloes. This is a potential limitation of the HEX technique, but if necessary dynamical friction could be added to the equation of motion. To do so would require an estimate of a subhalo's size and mass, information is not easily available until the simulation is post-processed by SUBFIND or unless some subhalo evolutionary model is assumed. Since the majority of our samples are small subhaloes of mass $\sim10^6M_{\odot}$, dynamical friction from both effects can therefore be discounted as a significant source of error in reproducing subhalo orbits. 

\par

The success of recreating orbits of subhaloes resimulated within a full HEX approximation is similar to that of single particles; most orbits are very well matched while others are not. We find that there is minimal difference between the orbits of subhaloes when treated as point masses and when treated as extended bodies. Over 99\% of subhaloes have a difference of less than 10\% (82\% less than 1\%) in their final energy when treated as point mass rather than as an extended object, and over 90\% have a difference of less than 10\% (43\% less than 1\%) in their final radial distance from the centre. This suggests that the extended nature of the subhalo has a minor effect on its motion, even though mass is being continuously stripped from the subhalo, forming leading and trailing streams.

\par

The cases where the Aquarius subhalo orbits and the resimulated orbits dramatically differ are again the result of encounter events. Subhaloes encounter the centre of the parent halo in the same way as particles, and any slight differences in the orbits are greatly amplified during the pericentric passage. However, as well as the passages near the centre, subhalo encounters are found to be more frequent than for single particles. When two subhaloes strongly interact, the orbit of at least one of the pair can be completely changed. In particular, a large subhalo merging into the parent halo will scatter any small subhaloes it passes as it falls in.  These subhalo-subhalo interactions are not well reproduced in the subhalo simulations using the HEX approximation since, while contributions to the potential from subhaloes are included, these are not well enough resolved with the number of basis functions we use to model them. Instead, the potential from subhaloes is blurred out.

\subsubsection{Evolution}

As subhaloes orbit within their parent halo they are tidally stripped
and shocked, losing mass and decreasing in size. Exactly how subhaloes
evolve and their final fate is a problem that has been extensively
studied \citep{Penarrubia:2005, Angulo:2009}. We resimulate subhaloes in three different potential
expansions corresponding to differing levels of sophistication. The
simplest is a fixed, spherically symmetric Hernquist potential, an
example of an analytical potential that is commonly used to represent
dark matter haloes in simulations \citep{Adams:2005,
Bullock:2005}. The second is a HEX potential that includes only
radial basis functions to obtain the correct radial mass distribution,
but with no information about the shape of the halo. The final
potential is a full HEX potential including both radial and angular
terms. We use the three different potentials in order to assess the
difference between the evolution of subhaloes using the commonly
employed method with a static simple potential and the effect of using a
full time-varying triaxial approximation.

\par

The parameters for the Hernquist potential are chosen so that it
matches the lowest order basis function from the expansion of the halo
at $z=0$. It has a scalelength of 33 kpc and a total mass of
$2\times10^{12}M_{\odot}$. This is a good fit to the halo at the final
time but overestimates the mass at earlier times. The second potential
(HEX$_{R}$), using only radial terms, has $n_{\rm max} = 20$ and
$l_{\rm max} = 0$, with a scalelength of 33 kpc and has time-varying
coefficients. The full potential (HEX$_{20}$), uses the default
parameters, so it has $n_{\rm max} = 20$ and $l_{\rm max} = 20$, is also
time-varying and has a scalelength of 33 kpc. Again, we exclude the
contribution to the HEX potential from the resimulated subhaloes. 

\begin{figure}
\centerline{\includegraphics[width=1.1\linewidth]{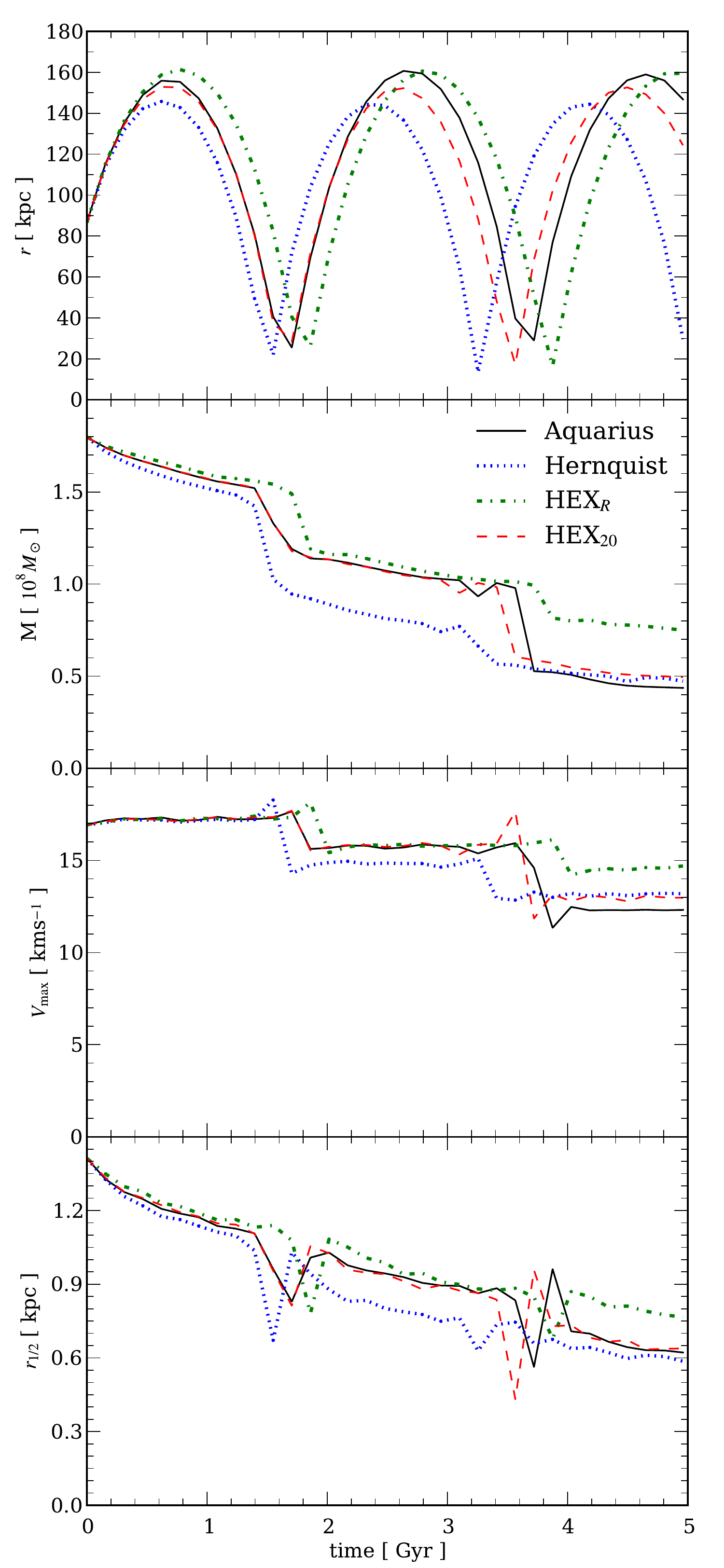}}
\caption{\label{fig:singleHaloProps} Comparison between the properties of different versions of the same subhalo. The full Aquarius Aq-A-2 simulation is represented by the black line. The other lines show resimulations of the subhalo in three differing potentials. \it Upper panel: \rm the distance of the subhalo from the centre of the parent halo. \it Upper middle panel: \rm the mass of the subhalo. \it Lower middle panel: \rm the maximum circular velocity. \it Bottom panel: \rm the half-mass radius.}
\end{figure}

\begin{figure}
\centerline{\includegraphics[width=1.1\linewidth]{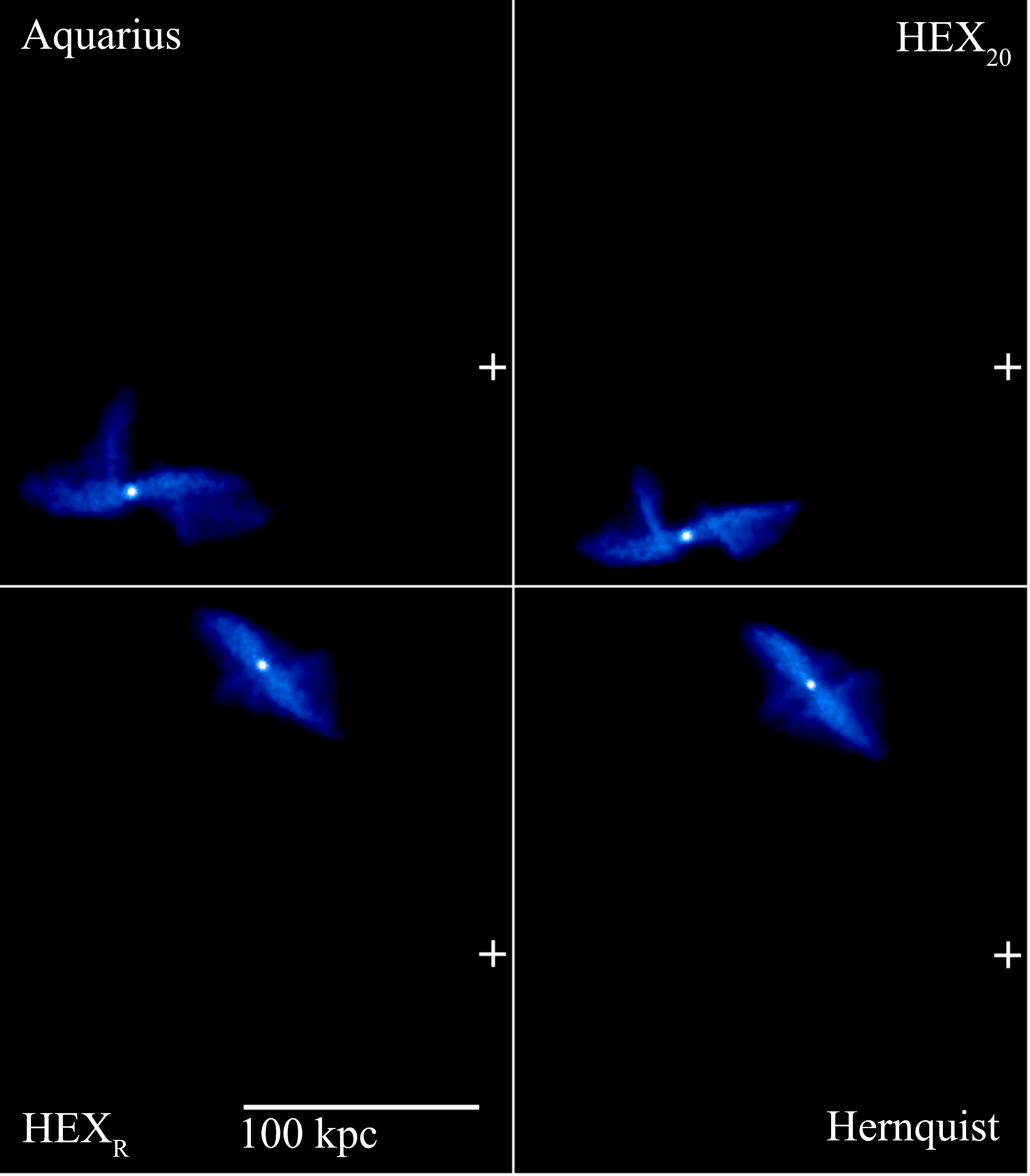}}
\caption{\label{fig:singleHaloImg} A projection of the smoothed density of a single subhalo resimulated in various different potential approximations at the subhalo's second apocentre. The subhalo reaches second apocentre at different times in the resimulations. The cross marks the centre of the parent halo in each case. \it Upper left panel: \rm the subhalo at 2.6 Gyrs in the original Aquarius simulation. \it Upper right panel: \rm the subhalo at 2.6 Gyrs in the full HEX$_{20}$ potential \it Lower left panel: \rm the subhalo at 2.8 Gyrs in the HEX$_R$ potential. \it Lower right panel: \rm at 2.3 Gyrs in a static Hernquist potential.}
\end{figure}

We start by focusing on a single subhalo to illustrate the technique in more detail. This subhalo has been selected from the Aq-A-2 simulation and contains 13120 particles, with a total mass of $1.8 \times 10^{8} M_{\odot}$. The subhalo was selected at redshift $z=0.5$, and resimulated for 5 Gyrs, with output snapshots every 155 Myrs. It is compared to the same subhalo extracted at the same times from the Aq-A-2 simulation.

\par

Fig. \ref{fig:singleHaloProps} shows the radial distance of the
subhalo from the centre of the potential and three main structural
properties that describe the state of a subhalo: the mass, the maximum
circular velocity and the half-mass radius. The properties of
the subhaloes in the two simplest methods, the Hernquist potential and
the HEX$_R$, immediately diverge from that of the Aquarius simulation,
as a consequence of the fact that they follow different orbits, as may
be seen in the top panel. These different orbits cause the subhalo to
experience different tidal stripping and, at pericentre, different
amounts of tidal shocking, resulting in incorrect estimates of the
structural properties. In the HEX$_{20}$ resimulation the subhalo
follows an orbit very closely matching the actual subhalo's orbit for
the first 2.5 Gyrs, until, following the first pericentric passage,
the orbits begin to diverge. Subsequently, the Aquarius subhalo
reaches a greater apocentric distance and falls back in slightly
later. Following this, near the halo centre, the small differences in
the paths are sufficiently large that during the second passage the
HEX$_{20}$ resimulated subhalo and the original Aquarius subhalo pass
the centre on opposite sides and depart in different radial
directions.

\par

During the initial period while the orbit of the subhalo in the
HEX$_{20}$ resimulation closely follows the fiducial Aquarius orbit,
the subhalo properties, the mass, half-mass radius and maximum
circular velocity, are reproduced extremely well. The subhalo is
stripped and distorted in the same manner as in the Aquarius
simulation. The subhalo continuously loses mass as it orbits within
the parent halo, with sudden and large decreases during pericentric
passages. Similarly, the maximum circular velocity, which is
determined by the mass in the inner regions of the subhalo, is
unaffected as mass is stripped from the outer edge. It is only when
the subhalo makes a close approach to the parent halo centre and is
tidally shocked and subject to maximum tidal stripping that the
internal structure of the subhalo is notably changed. This behaviour
is seen both in the Aquarius simulation and the HEX$_{20}$
resimulation and indicates that the important gravitational mechanisms
- tidal stripping and shocking, responsible for the evolution of a
subhalo - are equivalently modelled by the full HEX potential as they
are in the full simulation.
 
\par

An instantaneous picture of the subhalo during its second apocentre
can be seen in Fig. \ref{fig:singleHaloImg}. Rather than comparing the
subhalo at the same time, it is fairer to compare it at the same
position along the orbit as this removes any difference in orbital
phase. The resimulated subhalo in the HEX$_{20}$ potential is
strikingly similar to the original Aquraius subhalo. It is close to the
correct position, at the correct time and has very similar tidal
tails. This similarity includes the small perpendicular protrusion to
the left of the subhalo, which is a result of the end of the trailing
tidal tail being broken off during the apocentric turn-around. In
contrast, there is little resemblance between the subhalo in either
the Hernquist or the HEX$_R$ resimulation and the Aquarius original,
though there is a strong resemblance between the two simulations. Both potentials
are spherical, confining the subhalo to orbit in a plane, and thus the
two potentials generate similarly shaped orbits. However, there
is a large phase difference between the two. The Hernquist subhalo
reaches the second apocentre 290 Myrs before the Aquarius subhalo,
while the HEX$_R$ reaches second apocentre 140 Myrs after the Aquarius
subhalo.
\par

The final values of the mass, maximum circular velocity and half-mass
radius, are similar in the Hernquist and HEX$_{20}$ resimulations but
this is more a coincidence than the result of the subhalo having
the correct evolution in the Hernquist potential. While not completely
correct, the evolution of the subhalo is much closer to the real case
when the full HEX potential is used than when the simplified
potentials are used. This suggests that both the radial mass
distribution and the angular shape of the halo are important for
reproducing correct orbits, which is a prerequisite to achieve similar
evolution.  

\subsubsection{Population Evolution}

To assess whether the evolutionary mechanisms on subhaloes are the
same even though the orbits may not exactly match, we now consider a
population of subhaloes and look at the statistical match between a
set of Aquarius reference subhaloes and resimulations of them in the
three potentials. From the Aq-A-2 simulation we again use the set of
selected subhaloes with 100 or more particles that are within 90 kpc
of centre of the parent halo at $z=0.5$. The particles belonging to
these subhaloes are then tracked forward in time to follow the
subhaloes' evolution in the full simulation.  

\begin{figure}
\centerline{\includegraphics[width=1.1\linewidth]{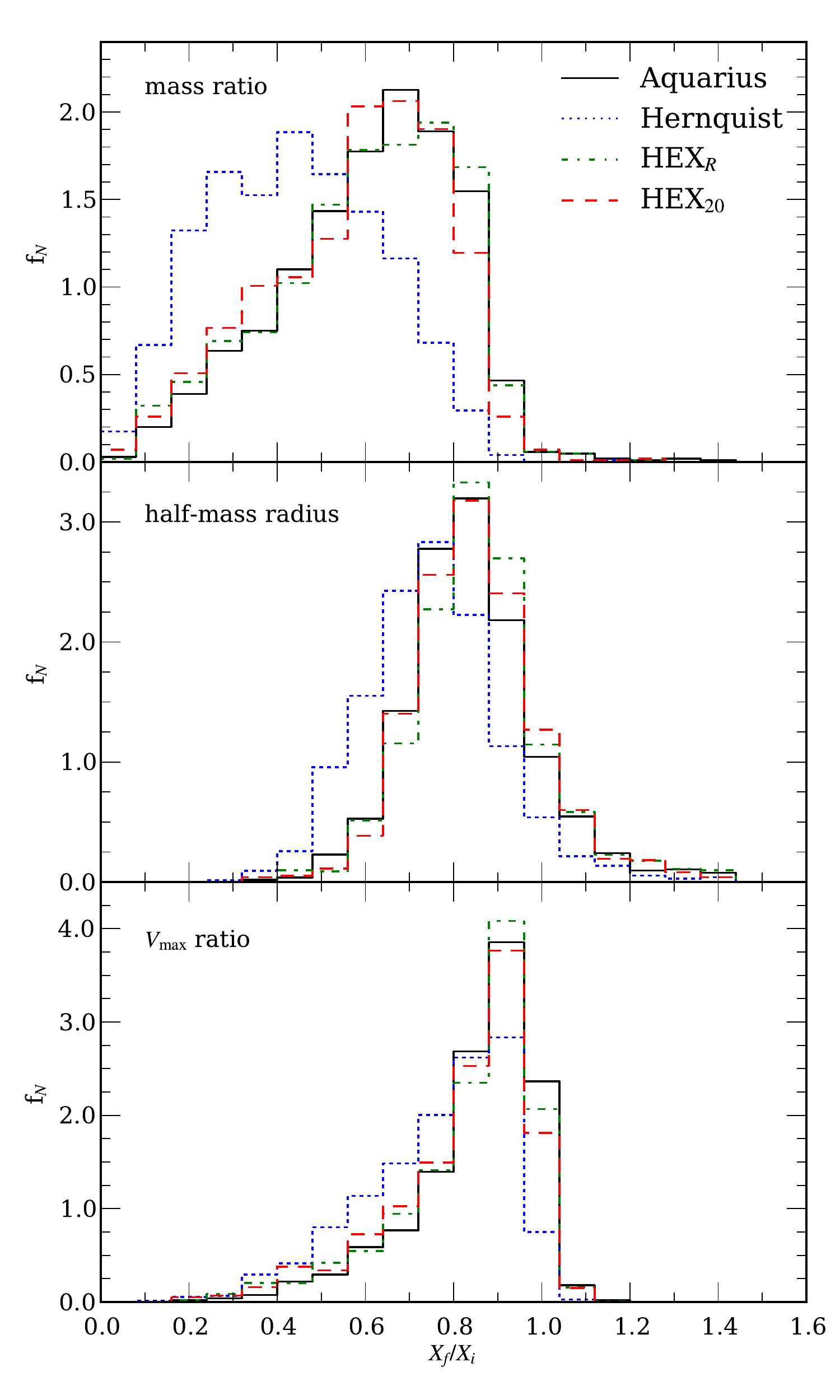}}
\caption{\label{fig:multiSubhaloProps} The distribution of X final
over X initial for a selected Aquarius subhalo population for three
different physical properties. The black line shows the actual
distribution that occurred in the original Aquarius simulation while
the other colours correspond to the different resimulations. For each
subhalo, the ratio is the given property at $z=0$ compared to its initial
value at $z=0.5$. \it Upper panel: \rm the distribution of final to
initial mass ratios. \it Middle panel: \rm the distribution 
of final and initial half-mass radii. \it
Bottom panel: \rm the distribution of final to initial maximum circular velocities.} 
\end{figure}

\par

Fig. \ref{fig:multiSubhaloProps} shows the population distribution of
the three main structural properties of subhaloes: the mass, the
half-mass radius, and the maximum circular velocity. The distribution
of the ratios of the final to the initial property has been used to
remove the influence of the property distribution and allow an easier
comparison of the actual evolution that the subhaloes undergo during 5
Gyrs. The distribution of mass ratios shows how much stripping the
subhaloes experience. Nearly all subhaloes in the Aquarius simulation
lose mass over the 5 Gyrs but a small fraction gain mass. The gain in
mass can be explained by inter-subhalo mergers, where two or more
subhaloes join to form a larger subhalo. The HEX resimulations and the
Aquarius simulation have the same small fraction of suhaloes
undergoing this mass increase; they have similar distributions of mass
ratios, with the same wide spread and a peak that occurs at 0.65. Only
the Hernquist potential shows significant difference.

\par

Similarly, the half-mass radius distribution is well matched by the
resimulations, except again by the Hernquist potential which is
slightly shifted to smaller sizes. Even though subhaloes generally
lose mass, a small proportion grow in size. This can occur when a
subhalo passes pericentre and is tidally shocked by the rapidly
changing potential field, thus increasing its internal energy and
resulting in an increase in size. This occurs in both the Aquarius
simulations and HEX resimulations. The maximum circular velocity
distribution is very slightly smaller in all the resimulations, with
the largest discrepancy again for the Hernquist population. The
primary reasons why the results from the Hernquist resimulation are so
different from the other two are the assumption of a static potential
of fixed mass throughout the whole simulation, which overestimates the
actual mass of the Aquarius halo at early times, and the fact that a
Hernquist potential gives the incorrect tidal radius for
subhaloes. The tidal radius is the distance from the centre of a
subhalo at which the gravitational tidal pull from the parent halo is
equal to the pull from the subhalo itself. Material outside of this
radius is stripped from the subhalo and becomes part of the parent
halo. We find that the Hernquist potential leads to underestimates of
the tidal radius for subhaloes that are between 30 and 200 kpc from
the centre of the parent halo and to overestimates outside this
range. The subhaloes therefore experience a different rate of
stripping over the course of their orbits than they do in the original
simulation and the other cases.

\par

Since the HEX$_R$ potential achieves an equally good match to the
Aquarius simulation as the full HEX potential that also includes the
angular terms, we conclude that the shape of the potential is
unimportant for reproducing the structural evolution of the subhalo
population in a statistical sense; only the radial mass distribution
needs to be correctly reproduced. The stripping of mass from a subhalo
is controlled by the tidal radius of the subhalo, so reproducing this
property correctly ensures the correct overall evolution. This can be
done by matching the radial mass distribution, which is easily
achieved with a small number of basis functions. In order to obtain
similar evolution on an individual subhalo basis, the orbits need to be
well matched, which does require the angular distribution and the full
HEX approximation.

\section{Application}

Having shown that the orbits, as well as the subhalo evolution, are
similar in a HEX approximation and in the original simulation, we now
demonstrate how the HEX technique can be used to go beyond the
original simulation. The introduction of new objects into the halo
that were not present in the original simulation, allows us to
investigate the reaction of these objects as if they had evolved in a
cosmologically realistic potential. They are unable to induce a back
reaction on the halo, but the method is appropriate for
studying light objects that would have had little effect on the
halo. This can be achieved at a much lower cost than re-running a
complete simulation and is more realistic than assuming a fixed
analytical profile, such as a Hernquist profile.
 
\subsection{Increasing Subhalo Resolution}

We now illustrate the technique of placing new, additional subhaloes
into the potential and simulating them at much higher resolution. As a
test, a subhalo is constructed to be similar to the subhaloes found in
the simulations, with an NFW density profile
\begin{equation}
\rho(r) = \frac{\rho_0}{\left(\frac{r}{r_s}\right)\left(1+\frac{r}{r_s}\right)^2},
\end{equation}
with $\rho_0 = 8\times10^7 M_{\odot} \mathrm{kpc}^{-3}$ and $r_{s} =
0.27$ kpc, and an isotropic velocity distribution. The subhlao is
injected into the HEX potential approximation of the Aq-A-2 halo. To
create equilibrium $N$-body halo realisations, we have used the
algorithm described in \citet{Kazantzidis:2004b} based on sampling the
phase-space distribution function to generate the subhalo. Since the
mass of an object with an NFW profile does not converge with radius,
we truncate the subhalo at the virial radius using an exponential
cut-off with a decay length set to ten times the virial radius. This
ensures the subhalo has a finite mass.

\par

We generate the initial subhalo at two resolutions. The first, lower
resolution version consists of 6000 particles with masses of
$1.4\times10^4 M_{\odot}$, the same particle mass as the Aq-A-2
simulation. The second version contains $10^6$ particles, a resolution
170 times higher, with individual particles masses of just $82
M_{\odot}$. Since the subhalo is small, with a SUBFIND mass of
$5\times10^7 M_{\odot}$, the absence of dynamical friction should not
be significant. The subhalo is placed 190 kpc from the halo centre,
approximately at the virial radius of the parent halo, where it will
be just entering into the main halo and will not yet have been
significantly stripped. The subhalo is simulated from $z=0.5$ for 5
Gyrs.
 
\par
\begin{figure}
\centerline{\includegraphics[width=1.1\linewidth]{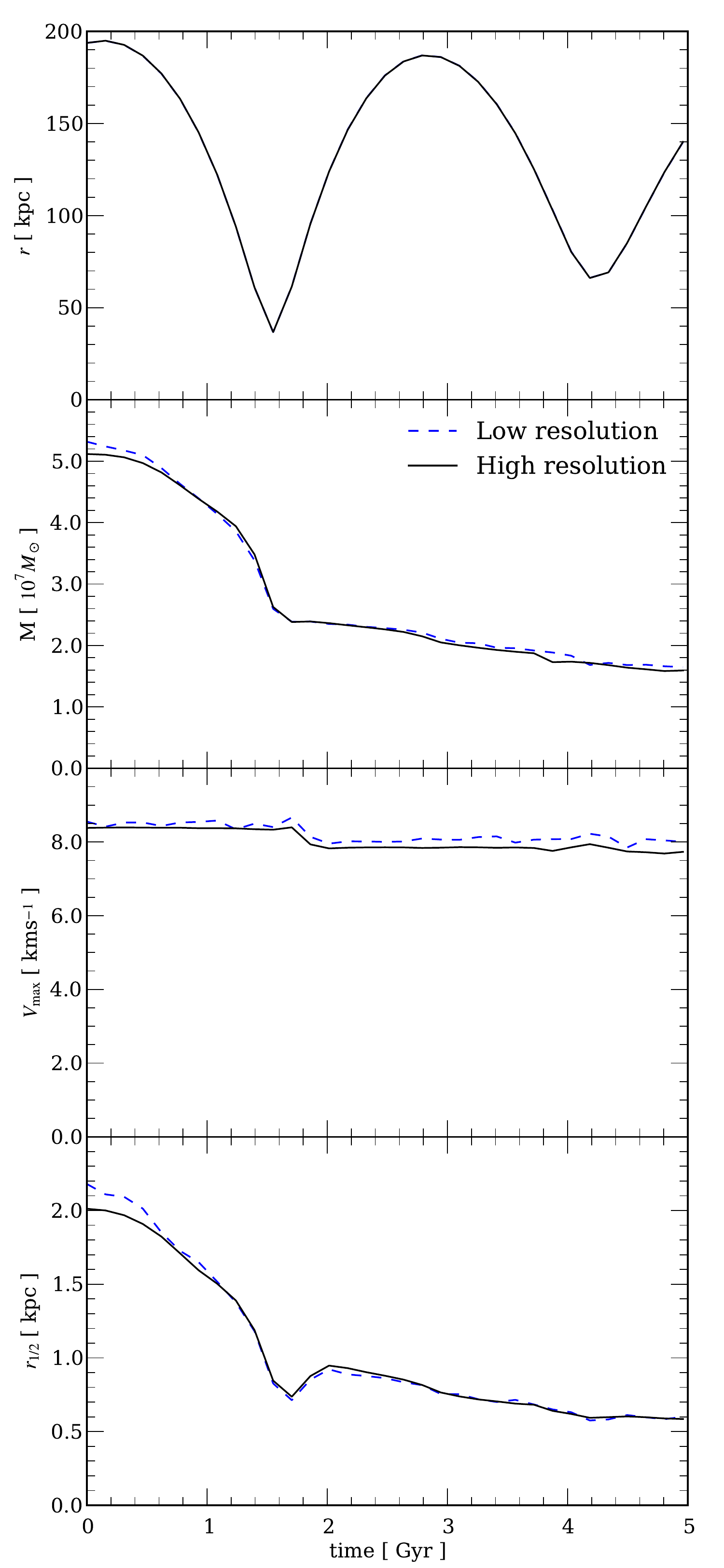}}
\caption{\label{fig:resolutionProps} Comparison between the properties given by SUBFIND for a subhalo simulated for 5 Gyrs at two resolutions in the HEX potential. \it Upper panel: \rm the distance of the subhalo from the centre of the parent halo. \it Upper middle panel: \rm the mass of the subhalo. \it Lower middle panel: \rm the maximum circular velocity. \it Bottom panel: \rm the half-mass radius.}
\end{figure}

The orbits of the two different resolution versions of the subhalo are
virtually identical. This is not unexpected, as we have already found
that subhaloes orbit as point masses regardless of their extended
nature. The changes in the properties of the subhalo over the 5 Gyr
simulation are shown in Fig. \ref{fig:resolutionProps}. Here we
compare the evolution of the mass, maximum
circular velocity and half-mass radius between the low and high resolution
simulations. While both realisations of the subhalo are sampled from
identical NFW profiles, the initial SUBFIND mass is slightly higher
for the low resolution version. Later mass estimates agree, suggesting
that in both cases the subhalo was stripped to the same tidal radius,
and the same material was lost regardless of whether SUBFIND had
initially associated it with the subhalo or not.

\par

The maximum circular velocities again are very slightly different, but
the higher resolution version has a smoother evolution since it is
less affected by noise from the discrete particle nature of the 
subhalo. The half-mass radius has the same initial discrepancy as the
mass, but again agrees at later times, with both versions undergoing
the same compression of the subhalo during the first pericentric
passage. Overall there is excellent convergence between the two
resolutions and it is clearly demonstrated that the structural
evolution is independent of the resolution of the subhalo as expected.

\par

\begin{figure}
\centerline{\includegraphics[width=1.1\linewidth]{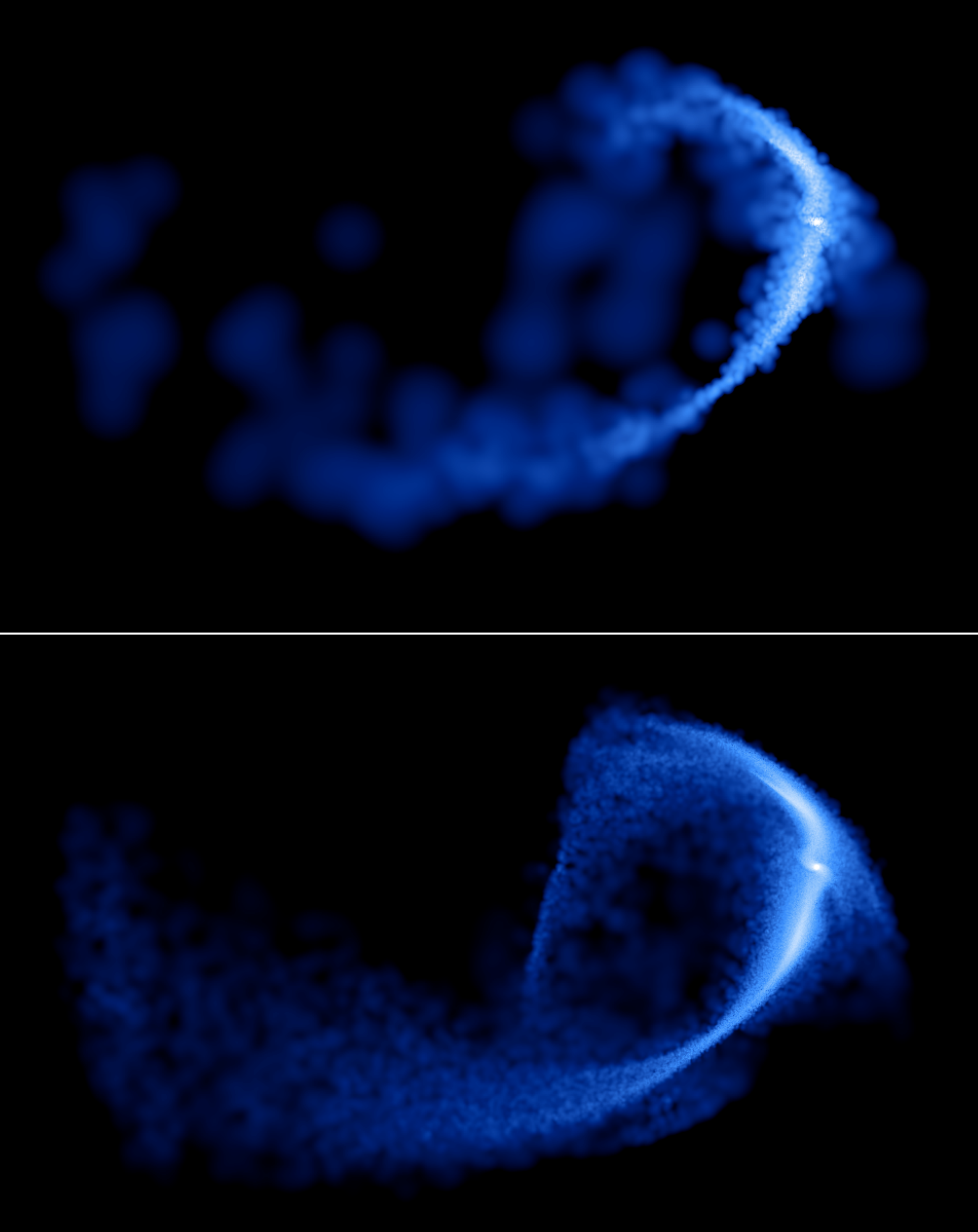}}
\caption{\label{fig:resolutionPicts} The smoothed density of the
resimulated subhalo after 5 Gyrs at $z=0$ using the HEX potential. \it
Upper picture: \rm the low resolution realisation subhalo containing
6000 particles. \it Lower picture: \rm the high resolution realisation
subhalo containing $10^6$ particles.}
\end{figure}

Apart from studying the subhalo we can compare the fate of the
material that is stripped from it and forms streams. There is both a
leading stream and a trailing stream, containing material that is no
longer bound to the subhalo but continues to follow similar
orbits. These streams match in the high and low resolution simulations
but are much clearer and can be traced much further in the high
resolution version. Sections of the streams containing a few tens of
particles in the low resolution version are now populated with
thousands of particles in the high resolution simulation. Features
that had been only hinted at are clearly defined in the high
resolution simulation. Especially clear are the caustics of the
streams which can be seen in Fig.~\ref{fig:resolutionPicts}. Another
feature that is not resolved in the low-resolution simulation but is
clearly visible in the high-resolution version is the bifurcation into
two separate arms of the leading tidal tail, the one above the subhalo
in Fig.~\ref{fig:resolutionPicts}.

\par

The HEX method allows us to simulate a subhalo at different
resolutions, with clear convergence between the two cases we have
examined. By focusing computing resources on just the subhalo and
using an approximation to the potential of the larger parent halo, we
have been able to reach an unprecedentedly high resolution, using a
particle mass of a few tens of solar masses and resolving tidal streams
much further and in a much sharper way than has been previously
achieved. The low-resolution simulation required only 15 cpu
hours\footnote{On a 2.2 GHz AMD Opteron (AMD Opteron 175)} and the
high-resolution subhalo only 2700 cpu hours. This is small compared to
the Aquarius A level 2 simulation, which has equivalent resolution to
the low-resolution subhalo and which took of order $\sim150,000$ cpu
hours over the same time interval. While a full simulation may include
thousands of subhaloes, we have demonstrated that is is possible to
vary the parameters and rerun multiple versions of a single subhalo in
a small fraction of the time.

%%%%%%%%%%%%%%%%%%%%%%%%%%%%%%%%%%%%%%

\section{Conclusions}

We have demonstrated the power of using the halo expansion method to
approximate a dark matter halo. While much work has previously been carried out using expansion methods as part of the SCF technique to calculate
the force in an N-body simulation, this is the first time that such an expansion technique has been
applied to describe an already simulated dark matter halo. Using a
small number of basis functions, the HEX technique offers a way to
approximate the time-evolving potential. A set of coefficients can be
calculated once from the simulation and then serve as a realistic
approximation of a halo. It is simple to integrate orbits within the
HEX potential approximation and, as a first test, we focused on
particle and subhalo orbits.

\par

Using the HEX method to represent a dark matter halo, however, has
some limitations. The potential is fixed and unable to react to
objects within it. New elements placed in the simulation, such as
additional subhaloes, cannot modify the halo potential. This could be
especially problematic when considering galaxies and the adiabatic
contraction that the presence of baryons is expected to produce. The
second major limitation is the lack of dynamical friction that should
be present in the equation of motions. Subhaloes orbiting within the
expansion are missing the effect of this force that would make their
orbits decay. While it is possible to add in dynamical friction
analytically, this requires assuming a model of subhalo evolution to
estimate the mass and size of the subhalo. 

\par

Through application of the HEX method to a halo simulated by the N-body code {\sc gadget}, we have demonstrated that:
\begin{itemize}
\item A HEX potential of a dark matter halo can approximate the halo
well enough to recover the radial component of the force to within 1\%
using only a few radial basis functions. 
\item  It is possible to integrate orbits within the expansions and
reproduce overall population trends. For individual orbits the degree
of success is varied. However, it must be remembered that {\sc gadget} dynamics are not necessarily numerically perfect and therefore differences are to be expected. For orbits that are near circular and stay
within the central 20 kpc of the halo we can accurately follow their
path over several dynamical timescales. 
\item Without dynamical friction subhaloes follow orbits close to those of point masses. Their extended nature and
tidal streams have little or no effect on their orbits. The orbits of
subhaloes are not simple planar orbits but involve complicated changes in
orientation and are strongly affected by encounters with the halo centre and other subhaloes.
\item The method can reproduce the structural evolution of individual
subhaloes. To obtain similar evolution for a particular subhalo we
need to match its orbit, which requires a full potential expansion. To
match the correct overall population evolution we do not need the full
expansion, but only the radial terms are required to obtain the
correct radial mass distribution. Not including the angular terms
greatly speeds up the force evaluation.
\end{itemize}

\par

We have been able to introduce new objects, such as subhaloes into the
HEX potential; we find an evolution consistent with that which would
have taken place if the subhaloes had been present in the original
Aquarius simulation. The technique allows us to simulate subhaloes
with much higher resolution than in the original simulation and
resolve features in the tidally stripped streams in great detail.

\par

While the HEX technique has some limitations it offers a powerful way
of improving current models of galaxy formation. The standard simple
spherically symmetric profiles often used to represent the dark matter
halo when modelling dynamical processes involving orbits miss
important effects related to the triaxiality of haloes and the
evolution of the potential. In order to build more realistic models it
is necessary, as we have shown, to use more sophisticated
representations of dark matter haloes such as the ones the HEX
technique offers. There is a large number of possible applications for
this technique and we have briefly explored only a few of these in
this paper.

\section*{Acknowledgments}

The simulations for the Aquarius Project were carried out at the
Leibniz Computing Centre, Garching, Germany, at the Computing Centre
of the Max-Planck-Society in Garching, at the Institute for
Computational Cosmology in Durham, and on the STELLA supercomputer of
the LOFAR experiment at the University of Groningen. BJL would like
to thank Andrew Cooper for useful discussions and suggestions. He is
supported by an STFC postgraduate studentship. CSF acknowledges a
Royal Society Wolfson Research Merit award. This work was supported in
part by an STFC rolling grant to the ICC. We would like to thank the referee 
for numerous useful comments that helped improve the paper.

\bibliographystyle{mn2e}
\bibliography{library}

\label{lastpage}
\end{document}